\documentclass[twocolumn]{aastex701}
\usepackage{xspace}
\usepackage{amsmath}
\usepackage{enumitem}

\newcommand{\ha}{H$\alpha$\xspace}
\newcommand{\hb}{H$\beta$\xspace}
\newcommand{\heii}{\ion{He}{2}\xspace}
\newcommand{\nii}{[\ion{N}{2}]\xspace}
\newcommand{\sii}{[\ion{S}{2}]\xspace}
\newcommand{\oii}{[\ion{O}{2}]\xspace}
\newcommand{\oiii}{[\ion{O}{3}]\xspace}

\newcommand{\niiu}{[\ion{N}{2}]$\lambda\lambda$6584\xspace}
\newcommand{\niilu}{[\ion{N}{2}]$\lambda\lambda$6548,84\xspace}
\newcommand{\siilu}{[\ion{S}{2}]$\lambda\lambda$6716,31\xspace}
\newcommand{\oiiilu}{[\ion{O}{3}]$\lambda\lambda$4959,5008\xspace}

\newcommand{\oiiihb}{[\ion{O}{3}]/H$\beta$\xspace}
\newcommand{\niiha}{[\ion{N}{2}]/H$\alpha$\xspace}
\newcommand{\siiha}{[\ion{S}{2}]/H$\alpha$\xspace}
\newcommand{\te}{$T_e$\xspace}
\newcommand{\tp}{$T_p$\xspace}
\newcommand{\tetp}{$T_e/T_p$\xspace}
\newcommand{\ibin}{$I_b/I_n$\xspace}
\newcommand{\ibincorr}{$I_b/I_n^{'}$\xspace}

\newcommand{\lyb}{Ly$\beta$\xspace}
\newcommand{\lyg}{Ly$\gamma$\xspace}

\newcommand{\kms}{km~s$^{-1}$\xspace}

\begin{document}
\title{SDSS-V LVM: Collisionless Shocks in the Supernova Remnant RCW86}
\author[0000-0002-6313-4597]{Sumit K. Sarbadhicary}
\affiliation{Department of Physics and Astronomy, The Johns Hopkins University, Baltimore, MD 21218 USA}
\affiliation{Department of Physics, The Ohio State University, Columbus, Ohio 43210, USA}
\affiliation{Center for Cosmology \& Astro-Particle Physics, The Ohio State University, Columbus, Ohio 43210, USA}
\email[show]{ssarbad1@jh.edu}

\author[0000-0002-4134-864X]{Knox S. Long}
\affiliation{Space Telescope Science Institute, 3700 San Martin Drive, Baltimore MD 21218, USA}
\altaffiliation{Visiting astronomer, Kitt Peak National Observatory}
\email{long@stsci.edu}

\author[0000-0002-7868-1622]{John C. Raymond}
\affiliation{Harvard-Smithsonian Center for Astrophysics, 60 Garden St., Cambridge, MA 02138, USA}
\email{jraymond@cfa.harvard.edu}

\author[0000-0001-8858-1943]{Ravi Sankrit}
\affiliation{Space Telescope Science Institute, 3700 San Martin Drive, Baltimore MD 21218, USA}
\email{rsankrit@stsci.edu}

\author[0000-0002-4755-118X]{Oleg V. Egorov}
\affiliation{Astronomisches Rechen-Institut, Zentrum f\"{u}r Astronomie der Universit\"{a}t Heidelberg, M\"{o}nchhofstra\ss e 12-14, D-69120 Heidelberg, Germany}
\email{oleg.egorov@uni-heidelberg.de}

\author[0000-0002-1379-4204]{Alexandre Roman-Lopes}
\affiliation{Department of Astronomy / Departamento de Astronomia, Universidad de La Serena, La Serena, Chile}
\email{aroman@userena.cl}

\author[0000-0003-4218-3944]{Guillermo A. Blanc}
\affiliation{Observatories of the Carnegie Institution for Science, 813 Santa Barbara Street, Pasadena, CA 91101, USA}
\affiliation{Departamento de Astronom\'{i}a, Universidad de Chile, Camino del Observatorio 1515, Las Condes, Santiago, Chile}
\email{gblancm@carnegiescience.edu}

\author[0000-0003-4679-1058]{Joseph D. Gelfand}
\affiliation{New York University Abu Dhabi, PO Box 129188, Abu Dhabi, United Arab Emirates}
\affiliation{Center for Astrophysics and Space Science (CASS), NYU Abu Dhabi, PO Box 129188, Abu Dhabi, United Arab Emirates}
\email{jg168@nyu.edu}

\author[0000-0003-3494-343X]{Carles Badenes}
\affiliation{Department of Physics and Astronomy and Pittsburgh Particle Physics, Astrophysics and Cosmology Center (PITT PACC), University of Pittsburgh, 3941 O’Hara Street, Pittsburgh, PA 15260, USA}
\email{badenes@pitt.edu}

\author[0000-0002-7339-3170]{Niv Drory}
\affiliation{McDonald Observatory, The University of Texas at Austin, 1 University Station, Austin, TX 78712-0259, USA}
\email{drory@astro.as.utexas.edu}

\author[0000-0003-3526-5052]{Jos\'e G. Fern\'andez-Trincado}
\affiliation{Instituto de Astronom\'ia, Universidad Cat\'olica del Norte, Av. Angamos 0610, Antofagasta, Chile}
\email{jose.fernandez@ucn.cl}

\author[0000-0002-8586-6721]{Pablo Garc\'{i}a}
\affiliation{Instituto de Astronom\'ia, Universidad Cat\'olica del Norte, Av. Angamos 0610, Antofagasta, Chile}
\affiliation{Chinese Academy of Sciences South America Center for Astronomy, National Astronomical Observatories, CAS, Beijing 100101, China}
\email{pablo.garcia@ucn.cl}

\author[0000-0002-2368-6469]{Evelyn J. Johnston}
\affiliation{Instituto de Estudios Astrof\'isicos, Facultad de Ingenier\'ia y Ciencias, Universidad Diego Portales, Av. Ej\'ercito Libertador 441, Santiago, Chile}
\email{evelynjohnston.astro@gmail.com}

\author[0000-0002-2262-8240]{Amy M. Jones}
\affiliation{Space Telescope Science Institute, 3700 San Martin Drive, Baltimore MD 21218, USA}
\email{amjones@stsci.edu}

\author[0000-0002-6425-6879]{Ivan Yu. Katkov}
\affiliation{New York University Abu Dhabi, PO Box 129188, Abu Dhabi, United Arab Emirates}
\affiliation{Center for Astrophysics and Space Science (CASS), NYU Abu Dhabi, PO Box 129188, Abu Dhabi, United Arab Emirates}
\affiliation{Sternberg Astronomical Institute, Lomonosov Moscow State University, Universitetskij pr., 13,  Moscow, 119234, Russia}
\email{ik52@nyu.edu}

\author[0000-0001-6551-3091]{Kathryn Kreckel}
\affiliation{Astronomisches Rechen-Institut, Zentrum f\"{u}r Astronomie der Universit\"{a}t Heidelberg, M\"{o}nchhofstra\ss e 12-14, D-69120 Heidelberg, Germany}
\email{kathryn.kreckel@uni-heidelberg.de}

\author[0000-0002-4825-9367]{Jing Li}
\affiliation{Astronomisches Rechen-Institut, Zentrum f\"{u}r Astronomie der Universit\"{a}t Heidelberg, M\"{o}nchhofstra\ss e 12-14, D-69120 Heidelberg, Germany}
\email{jing.li@uni-heidelberg.de}

\author[0000-0002-8931-2398]{Alfredo Mej\'ia-Narv\'aez}
\affiliation{Departamento de Astronom\'{i}a, Universidad de Chile, Camino del Observatorio 1515, Las Condes, Santiago, Chile}
\email{alfredo@das.uchile.cl}

\author[0000-0002-6972-6411]{J. Eduardo M\'endez-Delgado}
\affiliation{Instituto de Astronom\'{\i}a, Universidad Nacional Aut\'onoma de M\'exico, Ap. 70-264, 04510 CDMX, Mexico}
\email{jmendez@astro.unam.mx}

\author[0000-0003-2193-3005]{Rogelio Orozco-Duarte}
\affiliation{Instituto de Radioastronomía y Astrofísica. Universidad Nacional Autónoma de México, 58089 Morelia, Michoacán, México}
\email{rorozco@astro.unam.mx}

\author[0000-0001-6444-9307]{Sebastian Sanchez}
\affiliation{Instituto de Astronomía, Universidad Nacional Autónoma de México, A.P. 106, Ensenada 22800, BC, Mexico}
\email{sfsanchez@astro.unam.mx}

\author[0000-0002-7759-0585]{Tony Wong}
\affiliation{Department of Astronomy, University of Illinois, Urbana, IL 61801, USA}
\email{wongt@illinois.edu}

\collaboration{all}{SDSS-V LVM Collaboration}

\begin{abstract}

The supernova remnant (SNR) RCW86 is among the few SNRs with Balmer-emission lines containing broad and narrow spectral components that trace fast, non-radiative shocks in partially-ionized gas.\ These are invaluable laboratories for collisionless shock physics, especially for poorly-understood phenomena like electron-ion equilibration, and shock precursors. Here we present the first $\sim$0.3 pc spatial scale integral field unit (IFU) observations of the southwestern RCW86 shock, obtained as part of the Sloan Digital Sky Survey-V Local Volume Mapper (SDSS-V LVM). The forward shock, clearly visible as thin filaments in narrowband images, have broad H$\alpha$ components, indicating shock velocities varying from 500--900 km/s in the south to 1000--1500 km/s in the north. The varying velocity widths and broad-to-narrow intensity ratios show that electrons and ions have lower equilibration ($T_e/T_p \rightarrow 0.1$) in faster ($>$800 km/s) shocks, in line with previous studies. The broad components are generally redshifted from the narrow components by $\lesssim$100 km/s, likely due to shock-obliquity or non-Maxwellian post-shock distributions. We observe high extinction-corrected Balmer-decrements of 3--5 in the narrow components, indicating that conversion of Ly$\beta$ photons to H$\alpha$ is more efficient than Ly$\gamma$ to H$\beta$.  Broad HeII$\lambda$4686 was marginally ($\gtrsim$2$\sigma$) detected in the southern shock, meaning the shock is impacting gas with high ($>$30--100\%) neutral fraction. We also find the first evidence of an intermediate H$\alpha$ component in RCW86, with $\Delta$V(FWHM) = 193--207 km/s, likely due to a neutral precursor. We also briefly discuss the southwestern radiative shock, and lay out the exciting future of studying astrophysical shocks with LVM.
\end{abstract}

\section{Introduction}
\label{sec:intro}

Balmer-dominated supernova remnants (SNRs) are a rare group of optically-detected SNRs whose optical spectra are characterized by Balmer-emission lines (\ha, \hb) and little-to-no emission in the forbidden-line species of N, S, O etc, typically seen in radiative shocks \citep{Ghavamian2013}. Balmer-dominated shocks appear as thin ($\ll$ 1 pc) filaments tracing the non-radiative forward shock front. Most of the well-known examples of Balmer-dominated SNRs are young Type Ia SNRs like Tycho \citep{Kirshner1978, Smith1991, Ghavamian2000}, SN1006 \citep{Kirshner1987, Smith1991, Winkler1997}, Kepler \citep{Blair1991, Fesen1989}, RCW86 \citep{Long1990}, and the LMC SNRs N103B, 0509$-$67.5, 0519$-$69.0, DEM-L71, and 0548$-$70.4 \citep{Smith1991, Tuohy1982, Williams2014, Li2017, Ghavamian2017, Li2021}. Currently, the Cygnus Loop is the only evolved SNR of core-collapse origin where Balmer-dominated shocks have also been observed \citep{Raymond1983, Hester1994, medina14, Raymond2023a}.

The Balmer-emission manifests when fast (few$\times$100-1000 \kms) collisionless shocks propagate into a partially-ionized gas \citep{CR78, CKR80}. In the rapidly ionizing plasma behind such shocks, permitted lines are strongly favored over forbidden lines. However, before being ionized, the cold pre-shock neutral atoms (containing predominantly hydrogen) passing through the shock front undergo collisional excitations with the hot post-shock protons and electrons producing Balmer-line emission. The resulting line profiles have a characteristic narrow+broad shape, where the narrow component have widths $\sim$10 \kms that are proportional to the typical temperatures of the preshock warm neutral gas ($\sim$10$^4$ K), while the broad component ($\sim$10$^3$ \kms) is produced by the neutrals that undergo charge-exchange/transfer reactions with the hot post-shock protons, creating a `fast' neutral population with energies proportional to the high post-shock temperatures \citep[see Fig 1 in][for a schematic of a Balmer-dominated shock front]{Heng2010}.

These Balmer-dominated SNRs are invaluable laboratories for studying collisionless shocks \citep[e.g see][for summaries]{Raymond2001, Heng2010, Ghavamian2013, Raymond2023b}. Most astrophysical shocks in low-density plasma are \emph{collisionless}, i.e., the shock front is mediated by electromagnetic forces and turbulence in the plasma (as opposed to particle-particle collisions).\ Such shocks exhibit a number of interesting properties, such as non-Maxwellian velocity distributions, differential heating of electrons, ions and atoms, magnetic field amplification due to streaming instabilities, and cosmic-ray acceleration.\ Balmer-dominated shocks provide various observational tracers of the poorly-understood mechanisms driving these processes, particularly since the excitation of neutrals occurs in a thin layer ($\sim$10$^{15}$cm) in the post-shock ionization zone \citep{Heng07a, Heng07b}. For example, the line intensities and widths of the broad and narrow components have been widely used to study the temperature equilibration of electrons and ions in collisionless shocks \citep[e.g][]{CKR80, Smith1991, Ghavamian2001, Ghavamian2003, Rakowski2003, Ghavamian2007, VA08, Raymond2023b}. The line profiles and geometries of the narrow-line and broad-line shocks can also constrain heating from the hypothesized precursors formed by fast neutrals \citep[e.g][]{Blasi2012, Morlino2012} and cosmic-rays escaping upstream \citep[e.g][]{Raymond2011, Morlino2013}. When combined with proper motion measurements, the broad component can independently constrain the efficiency of cosmic-ray acceleration \citep[e.g][]{Morlino2013b,Morlino2013c,Morlino2014}, and provide an empirical distance measure to Galactic SNRs \citep{Winkler2003,Sankrit2005,VA08,Katsuda2008,Sankrit2016}

\begin{figure} 
    \includegraphics[width=\columnwidth]{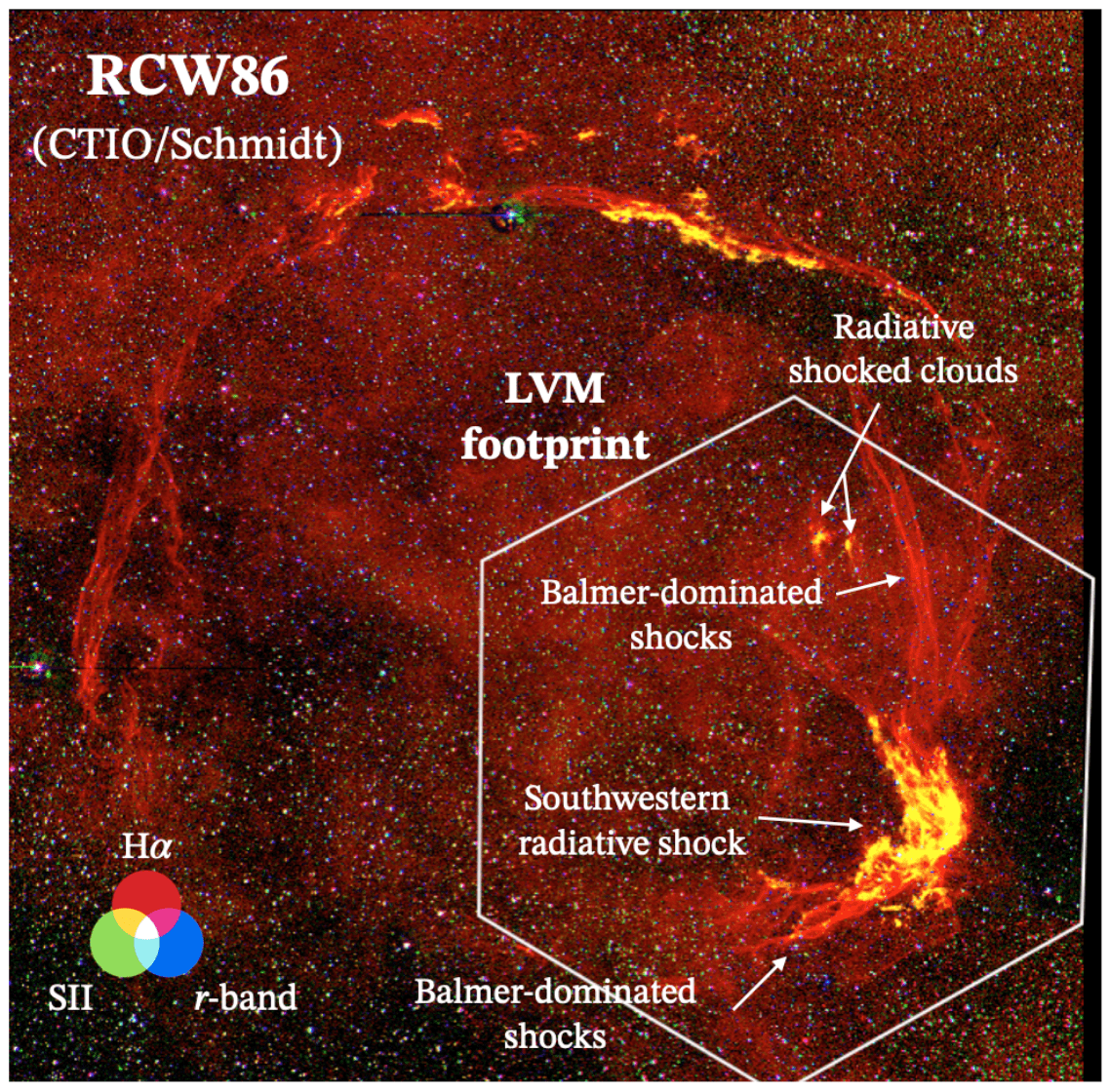}
    \caption{\textit{(Left)}: Narrowband optical image from CTIO/Schmidt (obtained by P. F. Winkler)  showing the full extent of the SNR RCW86. We use continuum-subtracted \ha (red), \sii (green) and an $r$-band image (blue) to produce a custom \emph{rgb} image that can distinguish the narrow Balmer-dominated filaments (mostly appearing red) from the radiative shocks (yellow). The observed hexagonal LVM footprint of the data presented in this work is shown in white.}
    \label{fig:lvmptg}
\end{figure}

In this paper, we present the first integral field unit (IFU) observations of one of the poster-child Balmer-dominated SNRs -- RCW86 -- conducted as part of the early-science verification phase with the recently commissioned Local Volume Mapper Instrument \citep[LVM-I][]{Konidaris2024}, an integral field unit (IFU) spectroscopic facility at Las Campanas Observatory, which is executing the Local Volume Mapper (LVM) program  \citep{LVM-Drory2024}, as part of the Sloan Digital Sky Survey V (Kollmeier, J.A., et al 2025, AJ, submitted). Widely believed to be the remnant of SN 185, a Type Ia SN in a wind-blow cavity \citep{Williams2011, Broersen2014}, RCW86 and its Balmer-dominated filaments were first reported by \cite{long90}, who used the width of the \ha\ profiles in the SW and W of RCW86 to estimate a shock velocity of 500-900 \kms. The extensive nature of these filaments across the RCW86 shock front was first reported by \cite{smith97} based on interference filter imagery (Figure \ref{fig:lvmptg}). Since then, various follow-up spectroscopy and multi-wavelength observations have been conducted on isolated portions of the shocks, leading to insights on cosmic-ray acceleration, electron-ion equilibration, and kinematic distance estimates \citep{Ghavamian2001, Sollerman2003, Helder2009, Helder2013, Morlino2014}. 



The paper is divided as follows -- Section \ref{sec:obs} describes the observations from the LVM exposures and data-reduction. Section \ref{sec:methods} describes our procedure for fitting the individual lines and their components with Gaussian models, and selection criteria with broad \ha. Section \ref{sec:results} presents the IFU maps of Balmer-lines in RCW86, in particular the spatial distribution of, and gas kinematics measured in, fibers with broad \ha. Section \ref{sec:nonradiativeshock} synthesizes these observations into key measurements relevant to the physics of collisionless shocks, such as electron-ion equilibration, Balmer-decrements affected by Lyman line trapping, He-line measurements and pre-shock neutral fraction, and intermediate \ha components from neutral precursors. Section \ref{sec:radiativeshock} provides a brief discussion of radiative shocks seen in the LVM pointing. 

\section{Observations} \label{sec:obs}

The LVM instrument comprises four separate 16~cm telescopes, including a science telescope feeding a fiber integral field spectrograph, covering the wavelength range 3600-9800 \AA\, with a spectral resolution at \ha\ of $R\sim4000$  or FWHM$\sim$75 km s$^{-1}$ \citep{LVM-Drory2024}.  Most of the fibers are dedicated to the science telescope IFU, which covers a field of about 25 arcmin in diameter with 1801 fiber-coupled $\sim$35.3 arcsec diameter spaxels \citep{LVM-Drory2024}.  Two of the other three telescopes are used for obtaining data for sky subtraction, and the final telescope is used to observe a number (up to 12) of F-type stars for calibration purposes.  For more details of the telescope and spectrograph design, we refer the reader to \cite{Konidaris2024, Herbst2024, Blanc2024} and Blanc et al (in prep).

The southwestern portion of RCW86, as shown in Figure \ref{fig:lvmptg}, was one of a small number of early science targets observed before the formal beginning of the survey. The region is only part of the larger SNR, which is about 25 pc in diameter \citep[assuming a distance of 2.5 kpc][]{Williams2011}, as can be seen in the narrowband optical image from the CTIO/Schmidt 0.4m telescope in Figure \ref{fig:lvmptg}. This specific pointing was chosen as it contains the prominent southwestern shocked cloud that typically features in images of RCW 86, and provides both prominent Balmer-dominated non-radiative shocks and forbidden-line dominated radiative shocks for our IFU study. Eventually, the entire SNR will be covered by the completed LVM survey, which will be analyzed in future work.

Specifically, four (undithered) 900 s exposures (exposure numbers 5059-5062, MJD 60203,60204), for a total of 3600~s, were obtained during this period. Each of these exposures reach 3$\sigma$ RMS sensitivities of $\sim$$(5.7-6.6) \times 10^{-15}$ erg cm$^{-2}$ s$^{-1}$ near the \ha line. For the purposes of our analysis, we begun with data products reduced to science-ready, sky-subtracted (\texttt{`SFrame'}) datasets using modules from version 1.1.1 of the LVM data-reduction pipeline \texttt{lvmdrp}  (Mej\'{i}a-Narv\'{a}ez et al., in prep).  These data products provide flux calibrated and nominally sky-subtracted spectra in each fiber, binned on a wavelength scale of 0.5 \AA.



\section{Methods} \label{sec:methods}

The primary goal in this paper is to understand the spatial distribution and physical properties of the Balmer-dominated shock front of RCW86 as revealed by the LVM data cube. We pursue this by fitting the bright Balmer lines \ha and \hb, and forbidden lines of interest such as \niilu, \siilu and \oiiilu, in order to extract the key observables -- the integrated line fluxes (erg s$^{-1}$ cm$^{-2}$), centroids (\AA), and full-width half maxima (FWHM, \kms) -- in each LVM fiber. 

Before carrying out the fitting, we perform two steps: 
\begin{enumerate}
\item We median-stacked the spectra from the four exposures fiber-by-fiber to reduce the noise. Offsets between exposures are too negligible to cause any issues with this stacking.
\item We subtracted  a ``local background'' spectrum in an attempt to remove any emission unrelated to the SNR that falls along the line of sight of the SNR. The ``local background'' fibers were selected by-eye from the pre-shock region, and exclude most of the SNR-emitting region. Details of this background subtraction is discussed in Appendix \ref{sec:backsub}, where we show examples of background-subtracted spectra, and also that our Balmer-dominated shock analysis is robust to the choice of background fibers. 
\end{enumerate}
\begin{figure*} \label{fig:ifupanel}
\centering
    \includegraphics[width=0.75\textwidth]{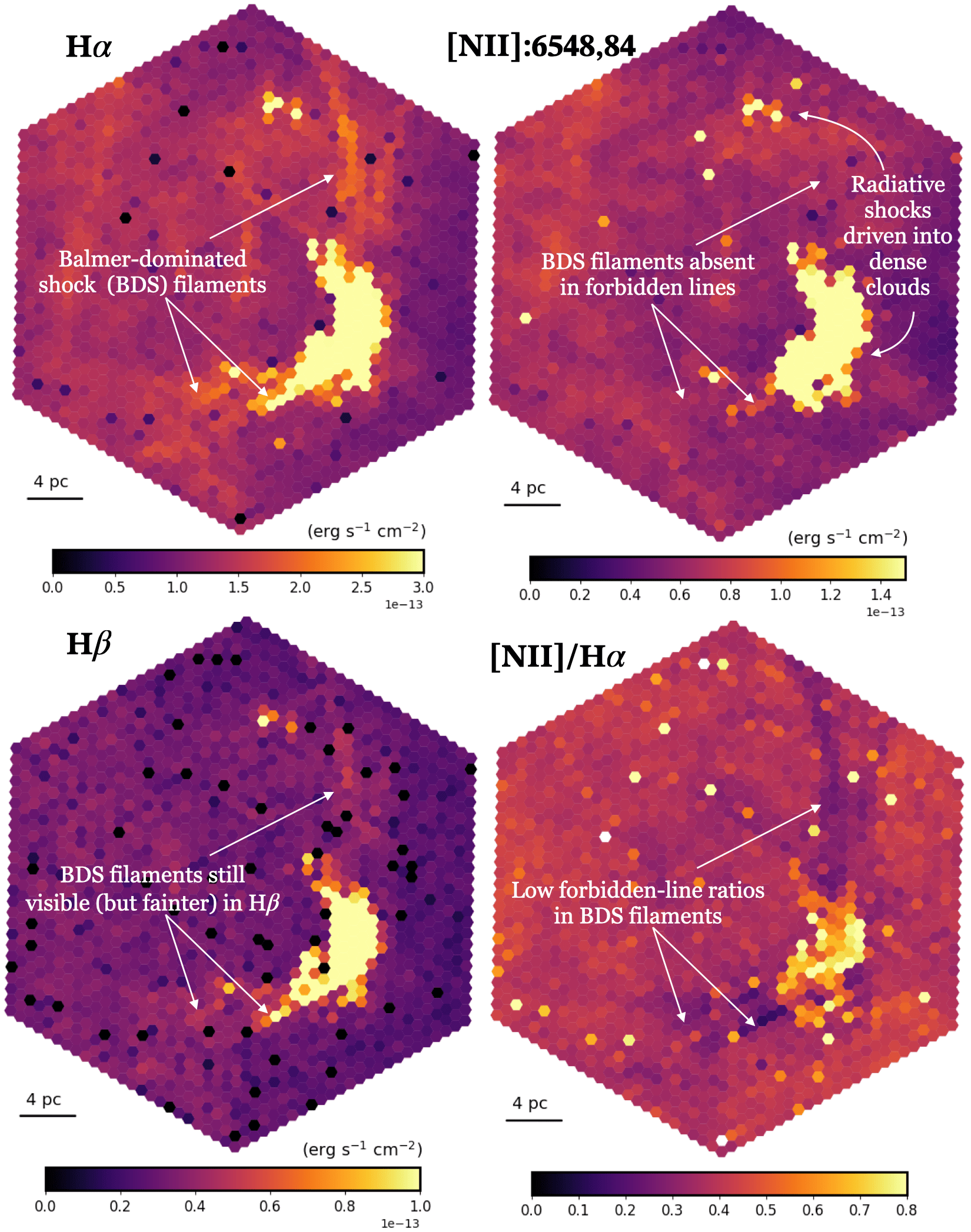}
    \caption{IFU maps of the southwestern region of RCW86 showing the integrated line fluxes of \ha and \hb, the prominent forbidden lines of \nii, and the ratio of \niiha from the narrow-only model. While it is more common to show [SII]:\ha ratios to indicate radiative shocks in SNRs, we show [NII]:\ha here, since [NII] is more relevant in the analysis that follows. Prominent Balmer-dominated filaments shown in Figure 1 are marked in each panel. The dense clouds pervaded by radiative shocks are bright and visible in all the maps, while they are particularly bright in the \niilu map. Note that the maps shown here are background un-subtracted, in order to show emission characteristics of both the SNR and the background.}
    \label{fig:ifumaps}
\end{figure*}

Within these median-stacked, background-subtracted IFU spectra, we custom-fit the individual line complexes of interest (e.g \ha + \nii) on a fiber-by-fiber basis with a multi-component model,
\begin{equation} \label{eq:flambda}
F_{\lambda} = \sum_{i} F_{\lambda, i} + F_c
\end{equation}
where $F_{\lambda,i}$ is the Gaussian fit to each emission line (or component) $i$, defined by 
\begin{equation}
F_{\lambda,i} = A_i \mathrm{exp}\left(-\frac{(\lambda-\mu_{i})^2}{2\sigma_{i}^2}\right)
\end{equation}
and $F_c$ fits the underlying stellar continuum, assumed to be linear over the finite wavelength range containing the lines
\begin{equation}
F_c = c_1 \lambda + c_2 \quad (\lambda_1 \leq \lambda \leq \lambda_2)
\end{equation}
where $c_1$, $c_2$ are constant coefficients. 

Analysis on the Balmer-dominated shock front in Sections \ref{sec:results} and \ref{sec:nonradiativeshock} will mostly focus on the \ha + \nii complex and \hb line, while the radiative shocks discussion in Section \ref{sec:radiativeshock} will use the fits to the \oiii and \sii lines. Since Balmer-dominated shocks are characterized by two-component (broad+narrow) \ha profiles, we fit the H$\alpha$+\ion{N}{2} complex in each fiber with two different models,
\begin{enumerate}
\item A \textbf{narrow-only H$\alpha$} model, which assumes the \ha has a single Gaussian component. The model consists of 11 parameters ($A_i$, $\mu_i$, $\sigma_i$ for [\ion{N}{2}]:6548,6584 and H$\alpha$, and $c_1$, $c_2$).
\item A \textbf{narrow+broad H$\alpha$} model, which assumes the \ha has a narrow and a broad component. This model contains 14 parameters (an additional $A_i$, $\mu_i$, $\sigma_i$ for the broad component of H$\alpha$). 
\end{enumerate}

Since the \nii-doublet properties are fixed by atomic physics and originating from the same gas, we fix their wavelengths and line-widths as, 
\begin{equation}
    \frac{\lambda_{NII:6548}}{\lambda_{NII:6584}} = \frac{6548}{6584},\     \frac{\sigma_{NII:6548}}{\sigma_{NII:6584}} = 1, 
\end{equation}
reducing the number of free parameters to 9 for the narrow-only model, and 11 for the narrow+broad model. We fit these models using the Levenberg-Marquardt Least-Squares fitting technique subroutines in the \texttt{astropy.modeling} package. The error bars of each parameter are estimated from the diagonal of the covariance matrix returned by the least-squares fitter.

From the fitted amplitudes and line-widths, we then measure the integrated line flux and full-width half-maxima (FWHM) of each line for our subsequent analyses as 
\begin{align}
I_{i} = \int F_{\lambda,i} \mathrm{d}\lambda = A_i \sigma_i \sqrt{2\pi}\\
\mathrm{FWHM}_{i} = 2.355\sigma_i.
\end{align}
 \begin{figure*}
     \centering
    \includegraphics[width=\textwidth]{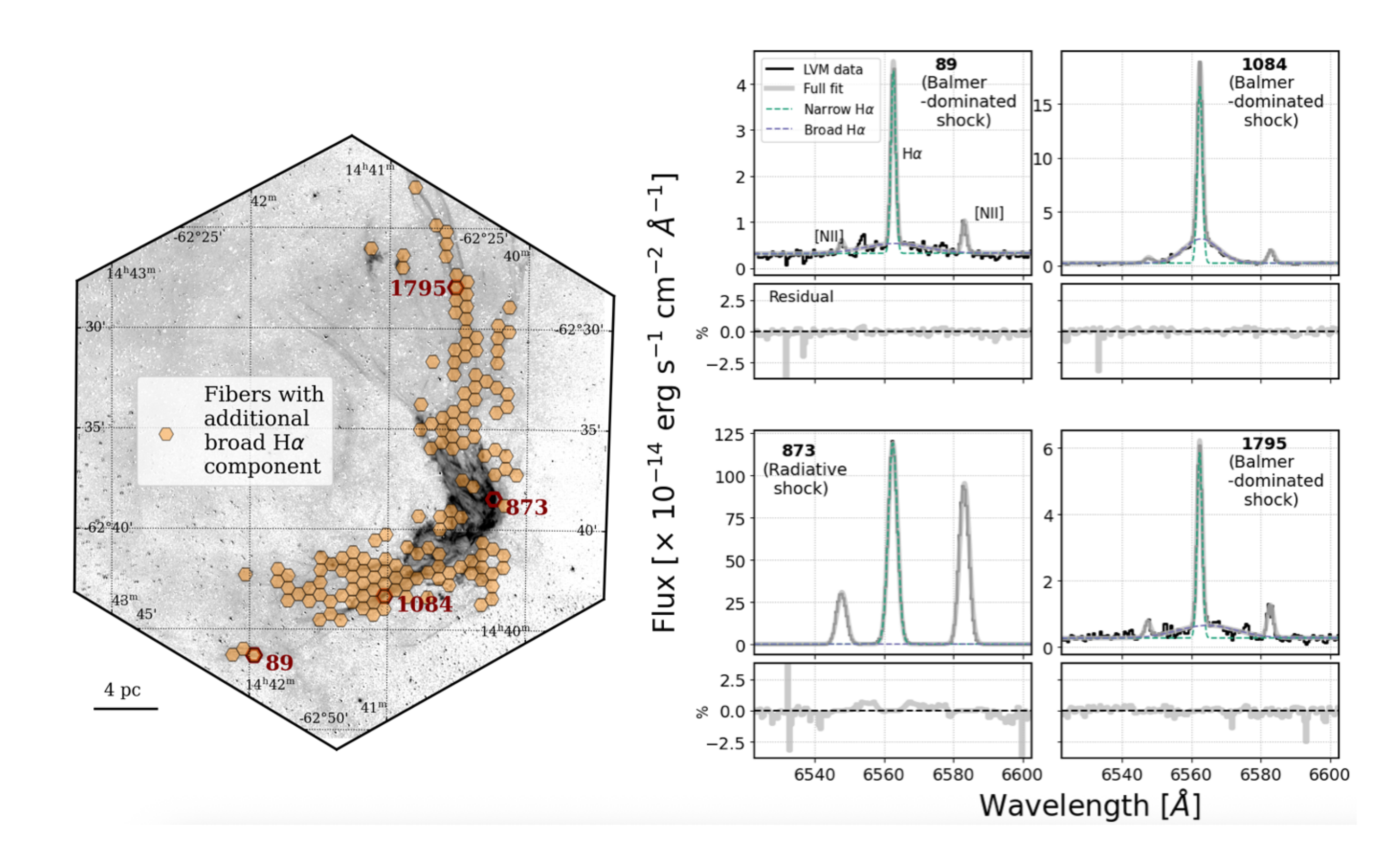}
    \caption{\textit{(Left)} Spatial distribution of fibers where an additional broad component was needed for the \ha line (orange hexagons; note  that fibers are circular) shown on the narrowband \ha image from Figure \ref{fig:lvmptg} Three such fibers with broad \ha (89,1084 and 1795) are labeled and shown on the right, along with one where no additional broad component was needed (873).  \textit{(Middle+Right panels)}: Spectra of the fibers from the left panel, showing the \ha+\nii complex, and the two-component Gaussian fit (broad in purple, narrow in green, their sum in thick gray lines). Residuals = (data$-$model)/data, expressed in percentages is shown in the smaller panel below each panel. The top two and lower right panels show fibers 89, 1084 and 1795 respectively, containing a clear excess broad emission, in addition to the narrow \ha fit -- characteristic of Balmer-dominated shocks. The bottom left panel shows fiber 873, containing a strong characteristic signature of a radiative shock, with no additional broad \ha component, and bright \nii lines.}
    \label{fig:bdsfront}
\end{figure*}

Error-bars of the intensities and FWHMs are estimated by error-propagation of the individual parameter errors.

We assume a given fiber is better fit by the narrow+broad \ha model over the narrow-only \ha model if it satisfies all of the following selection criteria.
\begin{enumerate}
    \item $\mathbf{\Delta \text{AIC} > 80}$ -- Here AIC refers to the Akaike Information Criteria \cite[see, e.g.][]{liddle07}, and is a quantitative method of model selection. For our purpose, the AIC is defined as\footnote{This is implemented in the \texttt{astropy.stats} package via the \texttt{akaike\_info\_criterion\_lsq()} module.} 
    \begin{align}
        \mathrm{AIC}= 2k - N \left(2\ \mathrm{ln}\frac{\mathrm{RSS}}{N}\right)\\
        \mathrm{RSS} = \sum_{j=0}^N \left( y_j - F_{\lambda,j}\right)^2
    \end{align}
    where $N$ is the number of data points, $k$ is the number of free-parameters in the model, and RSS is the residual sum of squares between the data points ($y_j$) and model fits to the data points ($F_{\lambda,j}$) defined in Eq \eqref{eq:flambda}. To determine whether the addition of a broad component improves the fit to the data, we use the difference in AIC between the narrow-only and narrow+broad models
    \begin{equation} \label{eq:aic}
        \Delta \mathrm{AIC} = \mathrm{AIC}_{\substack{\mathrm{narrow}\\\mathrm{only}}} - \mathrm{AIC}_{\substack{\mathrm{narrow}\\\mathrm{+broad}}}
    \end{equation}
     We chose the threshold of 80 to select as many fibers as possible with a genuine broad component, while minimizing the number of false positive detections (e.g broad components due to stars and/or a noisy continuum). See Appendix \ref{sec:badspectra} for details of this threshold choice.
    \item $\mathbf{A_i > 0}$, i.e fitted amplitudes of all the line components are positive. Line components in the \ha+\nii complex are negative in some fibers if there is a bright underlying stellar spectrum with absorption features. Small amounts of over-subtraction of the background in faint regions of the map can also cause this.
    \item $\mathbf{FWHM_{H\alpha, b}:FWHM_{H\alpha,n}>3}$, i.e ratio of the FWHM of the broad component is more than three times FWHM of the narrow component (see Appendix \ref{sec:badspectra} for details of this threshold choice). Recovered broad \ha being too narrow (e.g $<$200-300 km/s) particularly becomes an issue in fibers close to the radiative shock. While such narrow broad components is  theoretically possible in Balmer-dominated shocks (see Section \ref{sec:nonradiativeshock}), they become harder to distinguish from the broad and complex line profiles of recombining \ha in radiative shocks . 
    \item $\mathbf{FWHM_{H\alpha, b}:FWHM_{H\alpha,n}<20}$:\ This helps to exclude fibers with unusually broad components (we choose a threshold of 20 times the narrow FWHM). This happens in a minority of fibers dominated by stellar spectral features.
    \end{enumerate}
We refer the reader to Appendix \ref{sec:badspectra} and Figure \ref{fig:badspectra} for examples of fibers rejected by one or more of these criteria, justification of our chosen thresholds, and caveats involved. Note that this narrow+broad model, filtered through the above criteria, is only applied to the \ha+\nii complex for the purpose of identifying Balmer-dominated shocks. Analysis of the forbidden \oiii and \sii lines in Section \ref{sec:radiativeshock} assume single-Gaussian components.

\begin{figure*}
    \centering\
    \includegraphics[width=\textwidth]{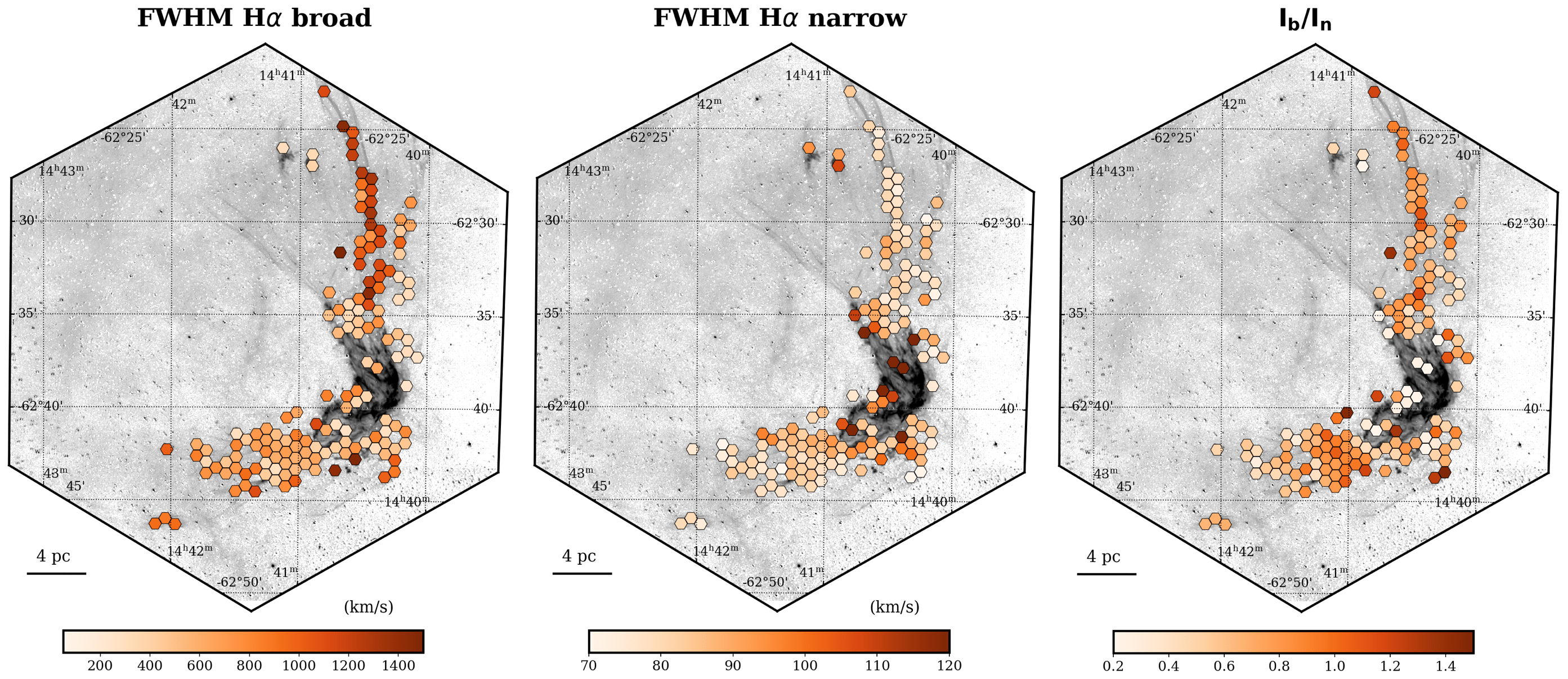}
    \caption{Physical properties of the 148 fibers with a detected broad component shown in Figure \ref{fig:bdsfront}. From left to right: FWHM of the broad \ha component, FWHM of the narrow \ha component, and \ibin ratio.}
    \label{fig:ifu-fwhm-ibin}
\end{figure*}

\section{Results} \label{sec:results}

In this section, we will first provide the IFU view of RCW86 obtained from the line-fitting procedure in Section \ref{sec:results:ifuview}, followed by a focused look at the distribution of the Balmer-dominated filaments in Section \ref{sec:results:bdsfront}. The physical properties of the shocks based on these measurements appear in Section \ref{sec:nonradiativeshock}.

\subsection{The optical IFU view of RCW86} \label{sec:results:ifuview}
Figure \ref{fig:ifumaps} shows the IFU maps of the Balmer lines \ha, \hb, and the prominent forbidden emission lines \niilu, obtained from the Narrow-only model fits. In these images, one can identify many of the spatial features seen in the narrowband CTIO image in Figure \ref{fig:lvmptg}, such as the Balmer-dominated shock filaments in the north and south, the prominent arc-shaped  southwestern radiative shocked cloud, and the smaller radiative shocked cloudlets in the north. The Balmer-dominated shock filaments are visible in the \ha (and to a fainter extent in the \hb image), but conspicuously absent or faint in the image of the forbidden lines \niilu. This is expected as Balmer-dominated shock trace the non-radiative forward shock, where post-shock temperatures are  too high ($\gtrsim$10$^7$ K) to allow cooling via forbidden lines. The morphologies of the shocked radiative clouds are prominent in the forbidden-line images. There is also diffuse emission within the IFU area detected in all the lines, also detected in the narrow band images shown in Figure \ref{fig:lvmptg} (the pervasive reddish hue; similar diffuse \sii is also present but suppressed in Figure \ref{fig:lvmptg} as we saturated the \ha more than \sii when making the figure). The emission extends beyond the forward shock of the SNR, so it is likely emitted by the intervening diffuse ISM. This however, does not affect our analysis of the shock front. The difference between the Balmer-dominated and radiative shocks are also visible in the line ratio map of \niiha shown in the fourth panel of Figure \ref{fig:ifumaps}. The radiative shocks have elevated line ratios of 0.6-0.8, while the Balmer-dominated shocks are clearly depressed in \niiha (with values in the range of 0.1-0.3). The curvature of the non-radiative forward shock can be clearly seen in the region showing depressed \niiha ratios. 

\subsection{Balmer-Dominated Shock front} \label{sec:results:bdsfront}
In this section, we discuss the spatial properties of the Balmer-dominated shock front inside the LVM footprint as traced by the presence of a broad \ha component.

Figure \ref{fig:bdsfront} shows the distribution of the 148 fibers where an additional broad component was identified by the Narrow+Broad gaussian fitting scheme described in Section \ref{sec:methods}. The pattern unmistakably traces out the curvature of the forward shock that can be seen as thin filaments in the IFU and narrowband \ha images in Figures \ref{fig:lvmptg} and \ref{fig:ifumaps}, and excludes most of the southwestern cloud that is primarily illuminated by the radiative shock front. Evidence of these broad features is shown in the spectra of the \ha and \nii complex in select fibers (Figure \ref{fig:ifumaps}). Fibers 89, 1084 and 1795 are among the fibers we consider part of the Balmer-dominated shock front and, as seen in Figure \ref{fig:bdsfront}, clearly contain an excess around the narrow \ha line that is better fitted with the inclusion of a broad \ha component. In contrast, the spectrum of the fiber 873 that squarely falls inside the radiative shocked cloud in the CTIO image does not show any such elevation, and is adequately fit with a single narrow component. Note that some radiative shock fibers do inevitably allow an additional broad component simply due to the flexibility of the broad+narrow model, but these fits are not significant based on their $\Delta$AIC values. A few more examples of such fits are shown and discussed in Appendix \ref{sec:badspectra}. We note that much of the \ha emission in the central and eastern part of the LVM field of view is consistent with having a single narrow component. This further supports the idea that this emission is likely just diffuse \ha emission along the line of sight, and not part of the forward shock seen face-on.

Figure \ref{fig:ifu-fwhm-ibin} shows the broad and narrow \ha FWHMs in the fibers identified as Balmer-dominated in Section \ref{sec:methods}. The narrow \ha FWHM have smaller values (70-90 \kms, close to the \ha line-spread function of $\sim$72 \kms) along the forward shock filaments in the north and south, and larger values ($\sim$100-140 \kms) closer to the shocked radiative cloud. The broad \ha FWHMs behave different from the narrow \ha, with the lowest values in the radiative shock, and highest along the Balmer-dominated filaments. This is because, as explained later in Section \ref{sec:epheating}, the broad FWHM is an order-of-unity tracer of the forward shock velocity, which is much higher than the radiative shock being driven into the dense southwestern cloud. The narrow \ha on the other hand, traces the cold neutrals, and is thus narrowest along the north and south Balmer-dominated filaments, and broadest near the radiative cloud, likely due to contamination with recombining \ha, with widths tracing the typical velocities of optically-visible radiative shocks \citep[$\gtrsim$100 \kms, e.g.][]{Allen2008, Long2022}. 

The broad \ha FWHM in Figure \ref{fig:ifu-fwhm-ibin} also suggests a gradient in the shock velocities of RCW86. The southern shock has FWHMs of $\sim$400-800 \kms, while the northern shock has FWHMs of $\sim$1000-1500 \kms. The slower  speeds in the southern part of the shock is qualitatively consistent with the ISM being slightly denser in the south, as seen from 21 cm atomic hydrogen observations \citep[e.g.][]{Ajello2016}. 

\section{Properties of the RCW86 Non-radiative shock} \label{sec:nonradiativeshock}
For the first time in RCW86, we have the opportunity to study the \emph{spatial} variation of the Balmer-dominated shock properties, and re-assess key observational diagnostics related to collisionless shock physics, particularly poorly-understood topics like the differential heating of electrons and ions by electromagnetic turbulence, pre-heating by shock precursors, and cosmic-rays. In the next few sections, we touch upon some of these major topics, though this is by no means an exhaustive investigation. 

\subsection{Electron-proton Heating} \label{sec:epheating}
The properties of the broad and narrow \ha components in Balmer-dominated shocks provide sensitive tests of electron-ion equilibration models in collisionless shocks. We refer the reader to the reviews of \cite{Heng2010} and \cite{Ghavamian2013} for the overview of the development of this area, and just provide here the salient features of the physics.

\subsubsection{Background}

As a result of being mediated by electromagnetic turbulence (unlike in collisional shocks), collisionless shocks can heat electrons to temperatures beyond their mass-proportional value ($T_e/T_p \gg m_e/m_p$), until they eventually equilibrate with the protons ($T_e/T_p \sim 1$) via Coulomb interactions. However, the timescales on which Coloumb interactions equilibrate the temperatures is much longer than ages of young SNRs like RCW86, making the \tetp vary with SNR age between these limits. The values of \tetp then depends on the numerous poorly-understood heating mechanisms in collisionless shocks, such as neutral, photoionization and cosmic-ray precursors, plasma wave heating and similar processes \citep[e.g][]{Vanthieghem2024}.

As noted earlier, in Balmer-dominated shocks the narrow component of the emission arises from neutral hydrogen atoms largely unaffected by the shock, which have not yet been ionized in the post-shock gas, while the broad component arises from a population of fast neutrals created by charge exchange.  As a result, the intensity ratio (\ibin) and FWHM of the broad and narrow \ha components provide a sensitive diagnostic for \tetp. The ratio is set by the balance between excitation, ionization, and charge exchange rates between neutrals and electrons, protons, and ions behind the shock -- which are strong functions of the shock velocity , \tp and \te, with some weak dependence on the pre-shock ionization fraction. The Balmer-emission also depends on the optical depth of the post-shock gas to Lyman line photons. Collisional excitation of the slower H atoms produces \ha and \lyb, the latter either escaping the gas under optically-thin conditions (Case A) or being re-absorbed by the gas (Case B) to further radiatively excite H atoms, leading to an enhancement of the narrow \ha.

Increasingly sophisticated models for obtaining \tetp from the broad and narrow components have been developed (and refined) over the years. The seminal works of \cite{CR78} and \cite{CKR80}, with subsequent corrections in \cite{Smith1991}, introduced the theory of two-component Balmer emission from shocks in partially-ionized media, dependencies on the excitation, ionization, charge transfer rates, and Lyman-line trapping. These models however tended to predict \ibin$>$1, which is inconsistent with observations. Later works \citep[e.g.][]{Ghavamian2001} expanded upon the originals by including more detailed treatments of the shock transition zone, and a Monte Carlo approach to the Lyman line trapping to produce Balmer photons. The works of \cite{Heng07a}, \cite{Heng07b} and \cite{VA08} further built upon these by accomodating for multiple charge transfer reactions per neutral atom. For this paper, we will comparing the predictions of the \cite{VA08} model with our measured \ibin data.


\subsubsection{Comparison between \ibin and the broad \ha FWHM} \label{sec:compareibinwithlit}

\begin{figure}
\includegraphics[width=\columnwidth]{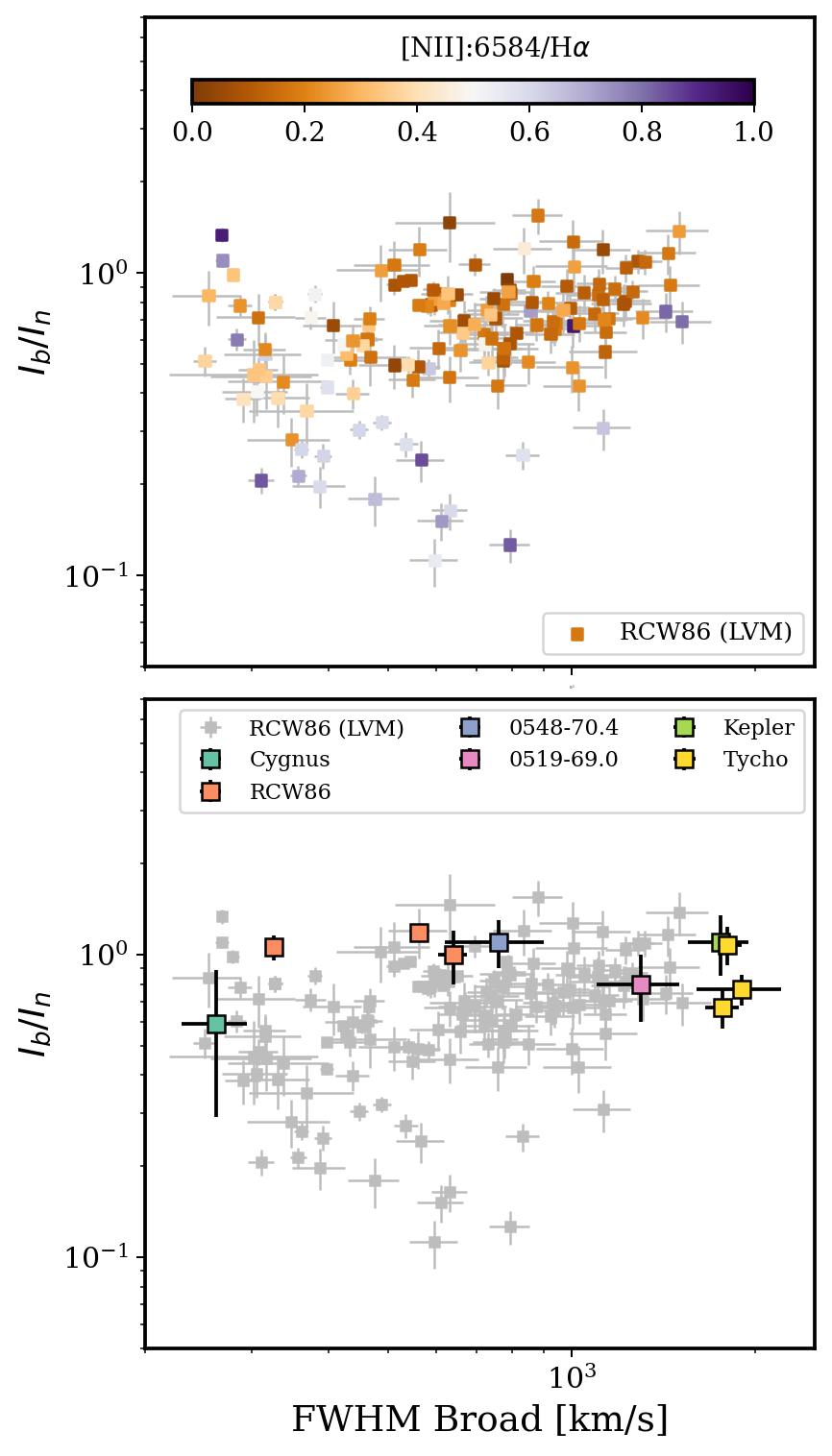}
\caption{$I_b/I_n$ vs full-width half-maxima of the 148 fibers with broad components detected in the \ha line. The top panel shows these points colored by their corresponding \niiha ratio. The bottom panel shows the same fibers but in grayscale, with similar estimates in isolated portions of different SNRs from the literature (compiled by van Adelsberg et al (2008)). }
\label{fig:ibin_fwhm_lit}
\end{figure}

Figure \ref{fig:ibin_fwhm_lit} (top panel) shows the relation between the \ha \ibin and FWHM of the broad component in RCW86. 
We only show fibers with a detected broad \ha component (Figure \ref{fig:bdsfront}) in Figure \ref{fig:ibin_fwhm_lit}. 
The \ibin-FWHM values span a wide range of values within this region of the SNR, with \ibin in the range of 0.1$-$2, and FWHM,
spanning $\sim$250$-$1500 \kms. A number of fibers with low \ibin ($<$0.3) have a strong radiative shock contamination, as evidenced by the \niiha ratios. Many of these fibers with low \ibin and high \niiha are coincident with the southwestern radiative shock as seen in Figure \ref{fig:ifu-fwhm-ibin}. They still passed our broad \ha selection criteria as they do indeed have a broad component, so it is likely that within the 35.3\arcsec fibers, we are seeing the superposition of strong radiative and Balmer-dominated filaments.
\begin{figure}
	\centering
	\includegraphics[width=\columnwidth]{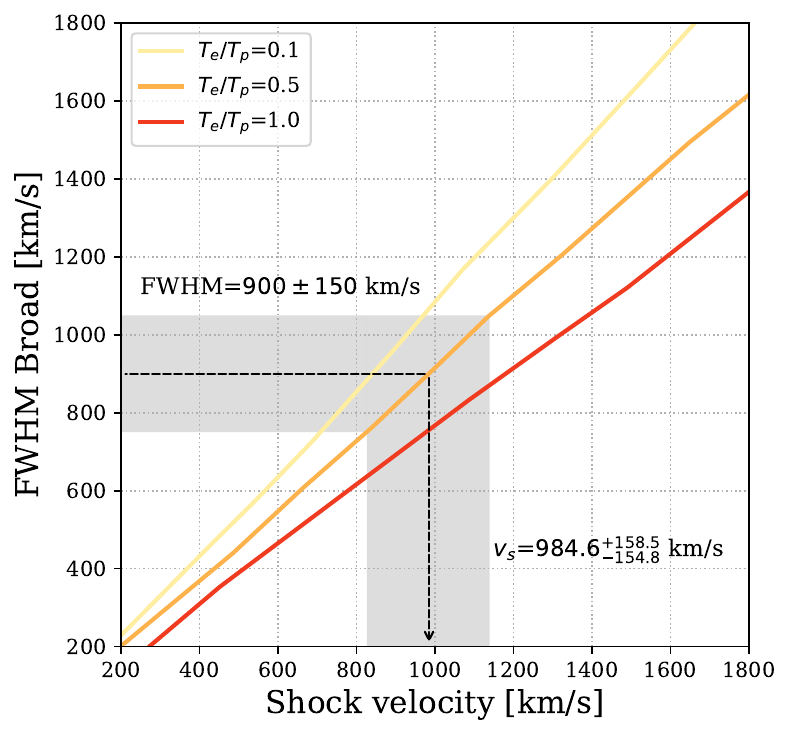}
	\caption{FWHM vs shock velocity for different values of \tetp predicted by the van Adelsberg et al (2008) models. We show an example of how we obtain shock velocity from a given value of FWHM (including uncertainties) for a particular \tetp (in this case, 0.5).}
	\label{fig:fwhm-vs-VA08}
\end{figure} 

We also compare the \ibin and broad \ha FWHM of RCW86 with other Balmer-dominated SNRs in the literature in the bottom panel of Figure \ref{fig:ibin_fwhm_lit}. We use the \ibin and FWHM values of SNRs compiled in Table 1 by \cite{VA08}, obtained from optical spectra in $\sim$few-arcsecond-wide slits, set on bright Balmer-dominated shock filaments in the SNRs. 
The impact of our IFU observations is immediately evident -- with a \emph{single} SNR, we are filling the parameter space of \ibin and FWHM with over a hundred measurements, covering almost an order of magnitude range in shock velocity.  This significantly increases the observational constraints on this parameter space that was previously only populated by a handful of isolated measurements from different SNRs. Figure \ref{fig:ibin_fwhm_lit} indicates that our measured \ibin values in the individual Balmer-dominated shock fibers of RCW86 are consistent, at least to order of unity, with previous measurements.
 

\subsubsection{Electron-proton equilibration from \ibin measurements} \label{subsec:tetp}
In this section, we check if the measured \ibin and broad \ha FWHM are consistent with predictions from collisionless shock models in VA08. Typically \ibin is compared with the shock velocity, which can be directly obtained from the FWHM. The relation between FWHM and shock velocity quantified in \cite{VA08} (their Figure 5) is shown in Figure \ref{fig:fwhm-vs-VA08}. Overall, the FWHM of the broad component is a useful order-of-unity measure of the shock velocity, as the former is set by the post-shock proton temperature which increases with velocity ($\propto v_s^2$). However, the proton temperature depends on \tetp. Higher \tetp means a larger share of the total energy is with the electrons, leading to lower proton temperatures and thus smaller FWHM at a given velocity. We use the relation to convert the FWHM measurements to shock velocity for a given value of \tetp (illustrated in Figure \ref{fig:fwhm-vs-VA08}).
\begin{figure*}
    \centering
    \includegraphics[width=\textwidth]{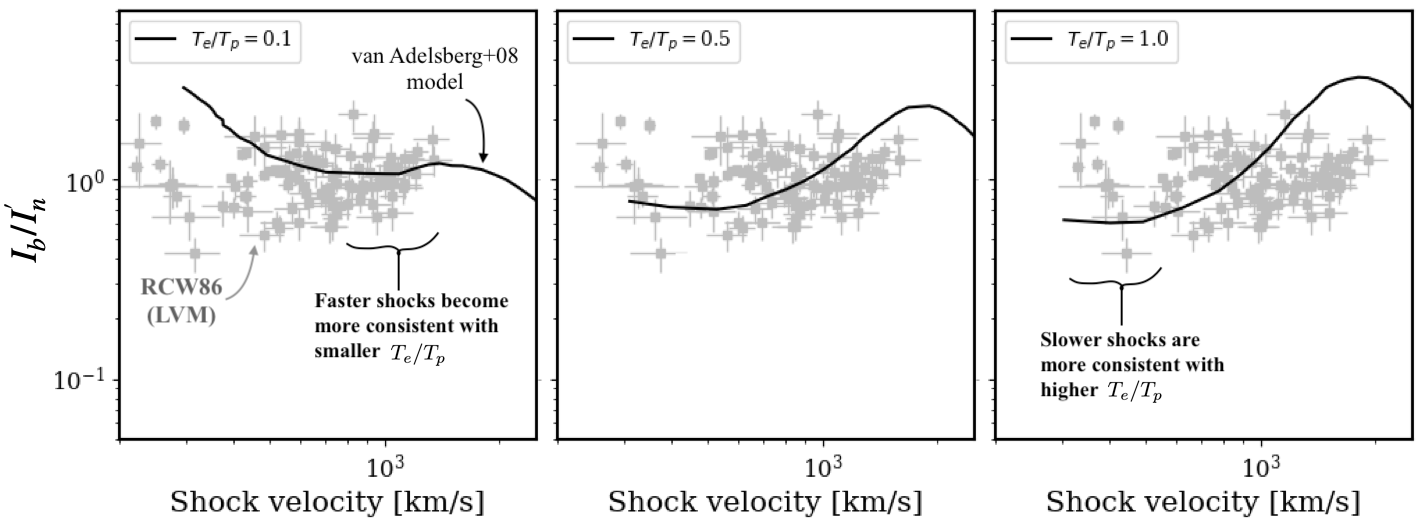}
    \caption{Comparison of our LVM data (gray) with the models of \cite{VA08} (black curved lines) for three values of \tetp. Each panel from left to right shows a particular equilibration model (\tetp). Since both the data and the models depend on \tetp (the shock velocities of the data are derived from the FWHM-\tetp relation in Figure \ref{fig:fwhm-vs-VA08}), we split them into the three panels. Data with \niiha$>$0.3 are excluded, and remaining data points have a fraction of narrow \ha flux removed due to suspected radiative shock contamination. The figure confirms that electrons and ions are not fully equilibrated in faster shocks, but tend towards full equilibration in slower shocks. See Section \ref{subsec:tetp} for details.}
    \label{fig:ibin-vs-VA08}
    \vspace{0.5cm}
\end{figure*}

Before comparing the observed \ha FWHM and \ibin with the \cite{VA08} models, we apply some corrections to the Balmer-line measurements to minimize  contamination from radiative shocks. We first exclude fibers with \niiha$>$0.3 in Figure \ref{fig:ibin_fwhm_lit} as the narrow \ha components in these fibers may be dominated by recombination. In the remaining fibers, there may still be some leftover contamination from recombining \ha in the observed $I_n$, since the LVM fibers are much larger ($\sim$35.3\arcsec) than the typical widths of Balmer-dominated shocks ($\ll$ 1 pc). As a result, the fibers could still include some radiative shock filaments behind the forward shocks \citep[see e.g.][]{Sankrit2023} or even from the surrounding diffuse ionized gas. We correct for this using the \niiu line intensity as a proxy by,
\begin{equation} \label{eq:radiativecorrectha}
    I_n^{'} = I_n - \dfrac{I_{\mathrm{NII},6584}}{\left(\mathrm{NII}/\mathrm{H}\alpha \right)_{\mathrm{rad}}}
\end{equation}
where $I_{\mathrm{NII,6584}}$ is the intensity of the fitted \niiu line, and $\left(\mathrm{NII}/\mathrm{H}\alpha \right)_{\mathrm{rad}}$ is the expected ratio for a purely radiative shock. As a first-pass at this subtraction, we assume this ratio to be 2/3, consistent with shock analyses of the Cygnus Loop \citep{Raymond2020}. Henceforth, we will be comparing $I_b/I_n^{'}$ with the VA08 models.

The variation of \ibin with shock velocity predicted by VA08 is shown in Figure \ref{fig:ibin-vs-VA08}. The figure is divided into three panels, with each panel corresponding to a specific \tetp model. The LVM data is also split into the three panels because the shock velocity is measured from the FWHM for a particular \tetp (Figure \ref{fig:fwhm-vs-VA08}). The VA08 models assume Case-B recombination of \lyb photons, a preshock ionization fraction of 50\%, and preshock He abundance of 10\% \citep[the variation of \ibin and FWHM with these parameters is negligible compared to $v_s$ and \tetp][]{Heng07a}. We do not show the Case A model as it predicts \ibin values much higher than most observations of Balmer-dominated shocks. The \ibin values are smaller for Case-B than Case A because the efficient absorption of \lyb photons leads to an increased flux of narrow \ha photons via radiative excitation \citep{CKR80}. 

The theoretical relation between \ibincorr and $v_s$ shown in Figure \ref{fig:ibin-vs-VA08}  is mainly set by the competition between rates of excitation, ionization and charge transfer of neutral atoms with electrons, protons, and ions \citep[e.g. Figure 3 in][]{Heng07a}. More excitations of neutral atoms lead to higher fluxes of narrow and broad \ha. More charge transfers specifically increase the production of fast neutrals (and thus, the  broad \ha flux), while more ionizations remove neutral atoms, reducing the narrow and broad \ha flux.  At low $v_s$, the charge transfer rates are substantially larger than ionization rate, which greatly increases the number of broad neutrals, leading to higher values of \ibincorr. At high $v_s$ ($>$2000 \kms), the opposite occurs, where the charge transfer rates rapidly decline with velocity, while the excitation and ionization rates increase monotonically, leading to decreasing \ibincorr. At around $v_s$$\sim$2000 \kms, a bump occurs in \ibin, corresponding to a maximum in the charge-transfer rates between ground and excited states, and a dip in the proton-atom and electron-atom ionization rates. As \tetp increases (going from left to right panel in Figure \ref{fig:ibin-vs-VA08}), the \ibincorr values at low $v_s$ decrease due to an increase in the electron ionization rate, while the bump at $v_s$ of 2000 \kms is higher, again due to more charge transfers than ionization.

The comparison of our data with the models in Figure \ref{fig:ibin-vs-VA08} indicates that electrons and protons have lower equilibration in faster shocks. The observed \ibincorr does not vary much with $v_s$, which can be attributed to a compensating variation on \tetp with the shock velocity. Shocks with speeds e.g. $>$1000 \kms become increasingly inconsistent with $T_e/T_p \rightarrow 1$ (third panel) and consistent with $T_e/T_p \rightarrow 0.1$. The opposite trend is seen as we move to lower shock velocities (e.g. $\lesssim$ 500 \kms), which are more coincident with $T_e/T_p \sim 0.5-1$. This is consistent with previous results that find \tetp $\propto$ $v_s^{-2}$ from inversion of the \ibin-$v_s$ and FWHM-$v_s$ relations for a grid of \tetp models \citep[e.g][]{Ghavamian2007, VA08, Raymond2023b}\footnote{We do not attempt a \tetp-$v_s$ analysis here since we do not have access to the full model grid ($m_e/m_p < T_e/T_p < 1$), so we cannot verify the relation with our data, but the result that \tetp decreases with $v_s$ is evident from our result.}. 
The relation has been interpreted as evidence that electron heating is nearly independent of velocity for $v_s \gtrsim 400$ \kms, possibly due to heating in a cosmic-ray precursor \citep[e.g][]{Ghavamian2007}. It is interesting to note the large scatter in the \ibin$-$$v_s$, likely due to varying conditions in the shocks, such as pre-shock densities, velocities, neutral-fractions, and radiative shock contributions. We defer a detailed investigation of these possibilities for future work, especially since \cite{VA08} does not include cosmic-ray or neutral precursors, which can be an additional source of heating that leads to lower \ibin values than predicted in Figure \ref{fig:ibin-vs-VA08} \citep{Morlino2012}. A more careful treatment of the conversion of \lyb photons to \ha is likely also needed, as current radiative transfer calculations do not account for the super-thermal broadening of the narrow \ha components ($\sim$30-50 \kms) often seen in Balmer-dominated SNRs \citep{Sollerman2003, medina14}.
\begin{figure*}
\centering
    \includegraphics[width=\textwidth]{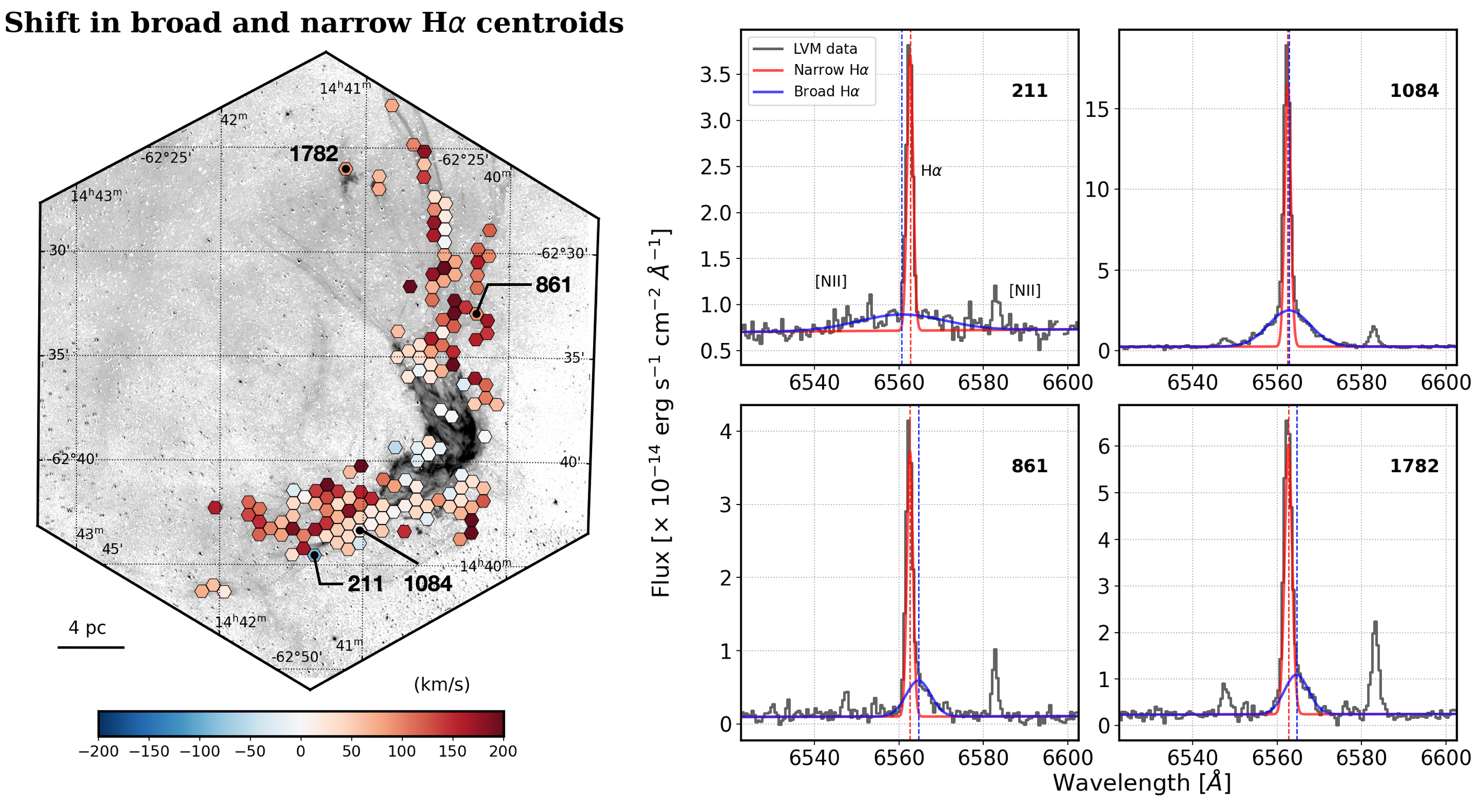}
    \caption{Relative radial velocity shifts between broad and narrow component centroids in the 148 fibers with a detected broad component. The shifts are given in velocity units as per Equation \eqref{eq:vshift}. Example \ha spectra of select fibers (indicated by arrow and numbers in the left image) are shown in the right. Red shows the narrow component while blue indicates the broad \ha component. The dashed vertical lines shown the centroid of the corresponding components. Fiber ids are given in the top right corner.}
    \label{fig:shift-broad-narrow}
    \vspace{0.5cm}
\end{figure*}
\subsection{Broad-Narrow \ha offset} \label{sec:shift-broad-narrow}

Our IFU data also provides the spatial variation in velocity offsets between the fitted centroids of the broad and narrow components of \ha. These offsets are plotted (in velocity units) in Figure \ref{fig:shift-broad-narrow}, quantified as:
\begin{equation} \label{eq:vshift}
    \Delta v_{\mathrm{shift}} = c\left(\frac{\lambda_{b}-\lambda_{n}}{\lambda_{n}}\right)
\end{equation}
where $\lambda_{n}$ and $\lambda_{b}$ are the fitted centroids to the narrow and broad components respectively. The offsets are typically redshifted, and increase as one moves towards the faster non-radiative shocks in the north and south. The northern Balmer-dominated shock show redshifted shocks of up to $\sim$100 \kms. The leading edge of the southern shock shows minimal offset, but the trailing part shows increased offsets between $+$100-200 \kms. Examples of the offset are shown in the spectra in the right panels of Figure \ref{fig:shift-broad-narrow}. Similar offsets have also been observed in other SNRs, such as Tycho \citep[490 \kms in the northeast,][]{Ghavamian2000, Heng07a} and N103B \citep[130-200 \kms,][]{Ghavamian2017}, which has been interpreted as resulting from the bulk motion of the post-shock gas into the plane of the sky. An alternative explanation is the projection of the velocity distribution of a non-Maxwellian population of post-shock ions \citep[see e.g.][]{Raymond2008, Raymond2010}.

\subsection{Balmer-decrements and Lyman-photon trapping} \label{sec:balmer-decrement}

The Balmer-dominated shock front is not only visible in \ha but also \hb as shown in Figure \ref{fig:ifumaps}. Although the filaments are clearly visible in \hb (indicated by arrows),
these are generally fainter than their \ha counterparts by a factor of 2.86. Higher values of Balmer-decrement are typically observed in interstellar plasma due to extinction, which causes light at bluer wavelengths to preferentially be absorbed and scattered. In Balmer-dominated shocks, the Balmer-decrement is also affected by Lyman-line trapping \citep{CKR80, Ghavamian2001}. Collisional excitation of H atoms to $n=1\rightarrow3$ results in \ha and \lyb photons, while $n=1\rightarrow4$ results in \lyg, \hb and Pa$\alpha$. Under Case-B conditions, where the post-shock plasma is optically thick to Lyman line photons, the \lyb and \lyg photons are efficiently re-absorbed by the post-shock gas, leading to radiatively excited \ha and \hb photons respectively. However, because the \lyb optical depth is larger than that of \lyg \citep{CKR80}, \hb is under-produced compared to \ha, leading to an enhancement in the \ha/\hb ratio beyond the expected theoretical value of 2.86. This enhancement in \ha/\hb is expected more in the narrow components than the broad, since the hotter neutrals are expected to experience fewer scatterings due to their large velocities. 
\begin{figure}
    \centering
    \includegraphics[width=\columnwidth]{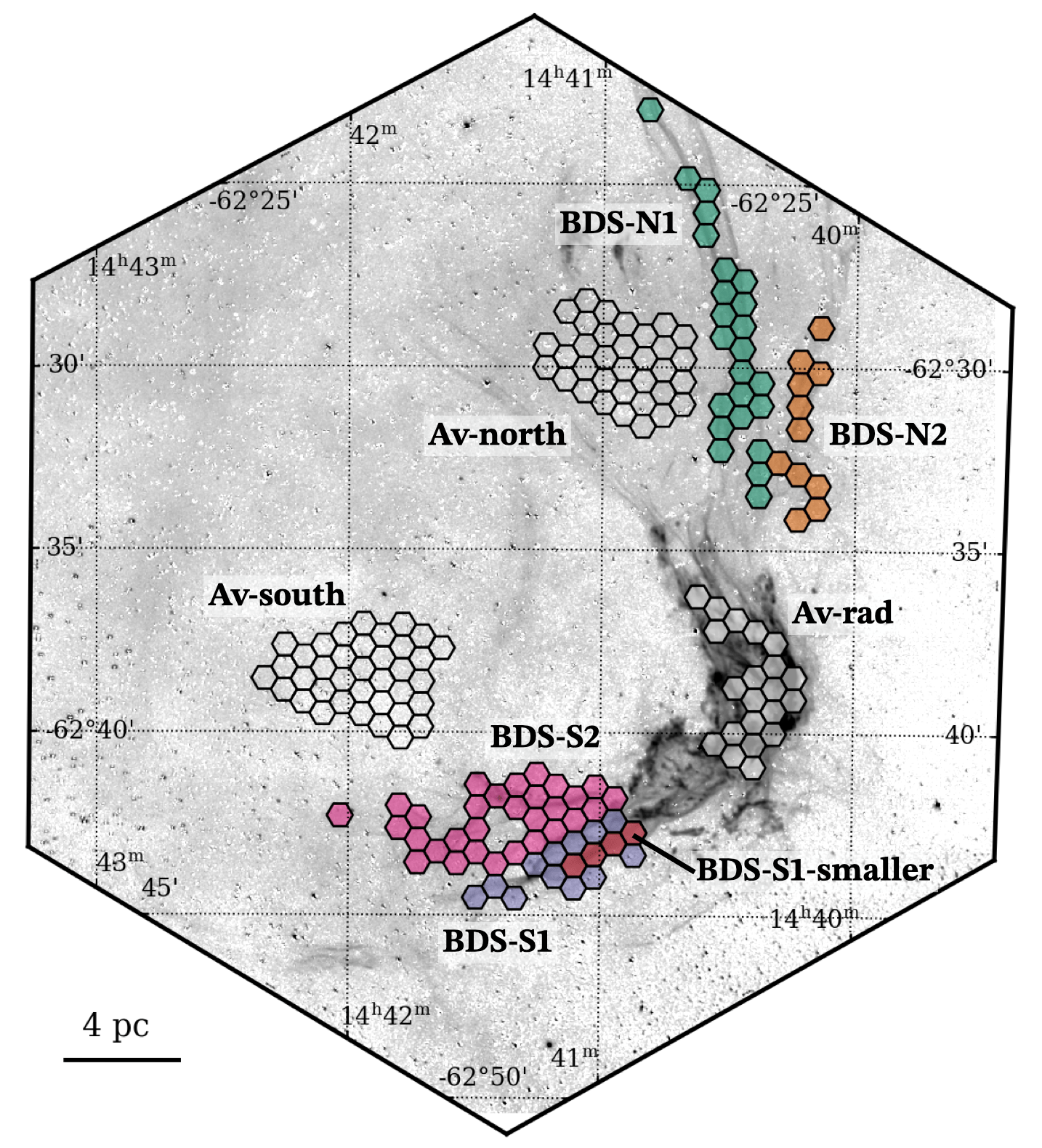}
    \caption{Regions where we stack the spectrum for analysis of fainter lines such as \hb and \ion{He}{1}, \ion{He}{2}, as discussed in Sections \ref{sec:balmer-decrement}. Fibers with the same color belong to one group (e.g BDS-N1=green, BDS-N2=gold, BDS-S2=magenta, BDS-S1=violet, and BDS-S1-smaller=red). Regions marked `Av-' are fibers we use for calculating extinction ($A_V$) for correcting the Balmer-decrements in the non-radiative shocks (see Section \ref{sec:balmer-decrement}).}
    \label{fig:stacked_regions_positions}
\end{figure}

To test this with our observations, we stack groups of Balmer-dominated fibers  to boost the signal-to-noise in the \hb line, to improve the measured properties of the fainter broad \hb line. These groups are shown in Figure \ref{fig:stacked_regions_positions}. We visually selected the fibers in each group from the 148 broad fibers that line up along prominent thin filaments seen in the narrowband image in Figure \ref{fig:lvmptg}, while avoiding going too deep into the bright radiative cloud (where the narrow component starts getting enhanced by recombining H). We divided them into two groups in the north (BDS-N1 and BDS-N2) and south (BDS-S1 and BDS-S2). We also include a smaller subset of BDS-S1 (BDS-S1-smaller), consisting of four fibers with the most prominent broad \ha components visually (a portion of these were also targeted in previous studies by \cite{Long1990} and \cite{Ghavamian2001}). These stacked regions will also be used later for analyzing He emission in Section \ref{sec:helines} and intermediate \ha components in Section \ref{sec:intermediateha}. For each group, we obtain a median-stacked spectrum by first shifting the spectrum in each fiber by the difference between their fitted narrow \ha centroid and the rest-frame \ha wavelength (6562.8 \AA). We then subtract the underlying continuum around the \ha and \hb lines for each spectrum (fitted with a 1D linear polynomial in the line-free parts), and obtain the 3$\sigma$-clipped median of the continuum-subtracted spectra of all fibers. With the median-stacked spectrum of each group, we fit the \ha and \hb lines with two component (narrow+broad) models to obtain intensities and FWHMs using the same method as described in Section \ref{sec:methods}. Also, just as we corrected \ha for contaminating \ha due to radiative shocks (Eq \ref{eq:radiativecorrectha}), we similarly correct the \hb line as follows
\begin{equation}
    I_{n,\mathrm{H\beta}}^{'} = I_{n,\mathrm{H\beta}} - \dfrac{I_{\mathrm{NII},6584}}{\left(\mathrm{NII}/\mathrm{H}\alpha \right)_{\mathrm{rad}}\left(\mathrm{H\alpha/H\beta}\right)_0}
\end{equation}
where $\left(\mathrm{NII}/\mathrm{H}\alpha \right)_{\mathrm{rad}}$ is again the expected ratio in radiative shocks ($=2/3$) and $\left(\mathrm{H\alpha}/\mathrm{H}\beta \right)_0 = 2.86$.

\begin{figure*}
    \centering
    \includegraphics[width=\textwidth]{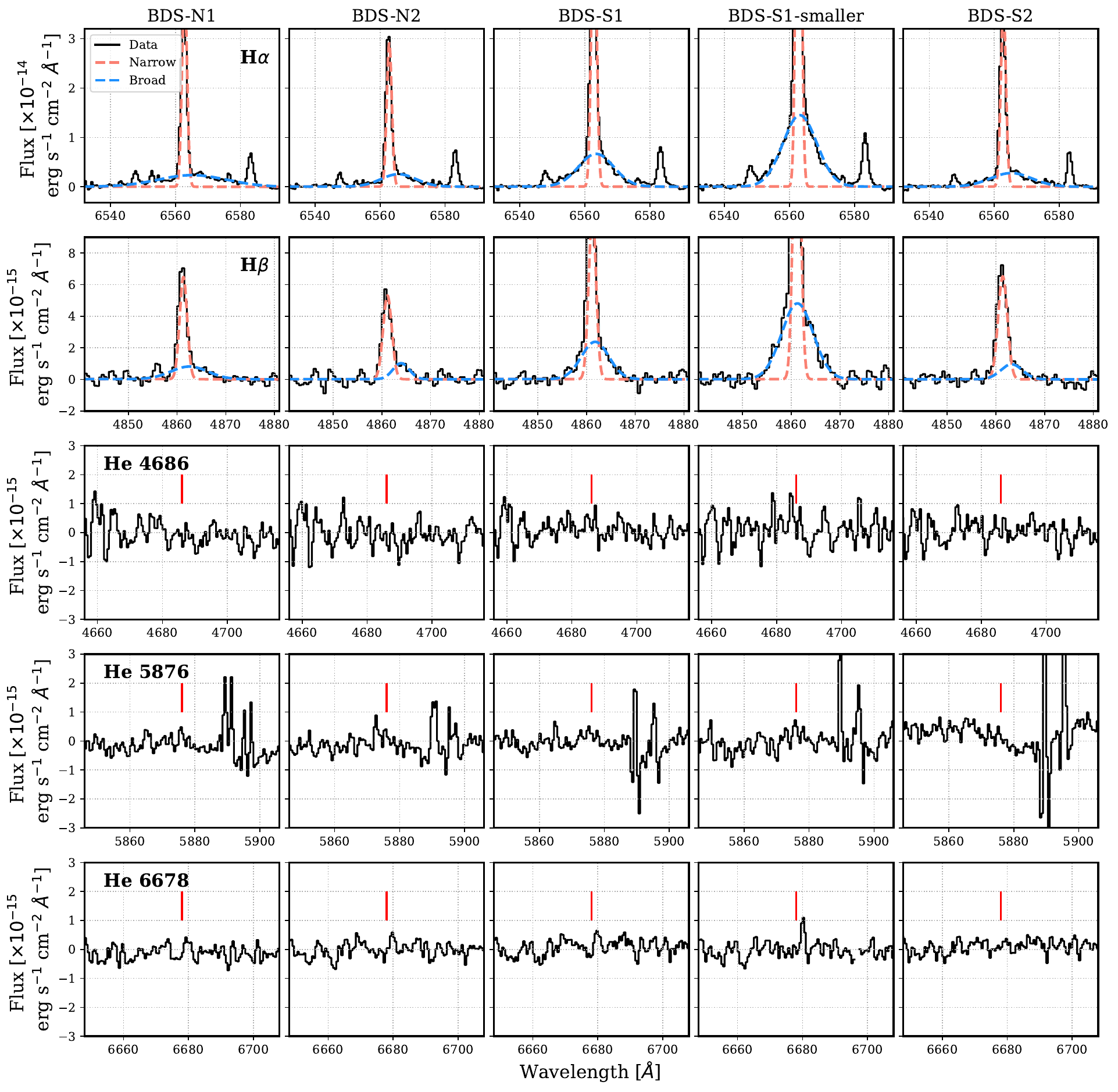}
    \caption{\ha and \hb profiles in the five stacked regions from Figure \ref{fig:stacked_regions_positions}, shown at the top of each panel. Black is the observed spectra (continuum-subtracted and shifted to rest-wavelengths of respective lines, see Section \ref{sec:balmer-decrement}), while red and blue are the best-fit narrow and broad components respectively. Bottom three rows show zoom-ins of the \ion{He}{2}$\lambda\lambda$ 4686, \ion{He}{1}$\lambda\lambda$5876 and \ion{He}{1}$\lambda\lambda$6678 lines (rest-wavelengths indicated by red vertical line).}
    \label{fig:stacked_regions_spectra}
\end{figure*}

Profiles of the broad and narrow component fits to the stacked \ha and \hb in each region are shown in Figure \ref{fig:stacked_regions_spectra}. Broad components can be seen in both \ha and \hb lines, further confirming that the fast neutrals emit across the Balmer-spectrum\footnote{We also detected narrow lines corresponding to the higher Balmer transitions such as H$\gamma$, H$\delta$, but attempted no identification of broad components due to their faintness. The \ha and \hb lines provide sufficient characterization of the Balmer-dominated shocks.}. In general, we find that the \ha and \hb profile shapes are correlated for the same region. BDS-S1 and BDS-S1-smaller have bright and symmetric broad components in both \ha and \hb. Offsets between the broad and narrow components that were prominent in the individual fibers of BDS-N2 and BDS-S2 (Figure \ref{fig:shift-broad-narrow}, Section \ref{sec:shift-broad-narrow}) are also apparent in the integrated profiles of \ha and \hb. 

\begin{figure*}
    \centering
    \includegraphics[width=0.9\textwidth]{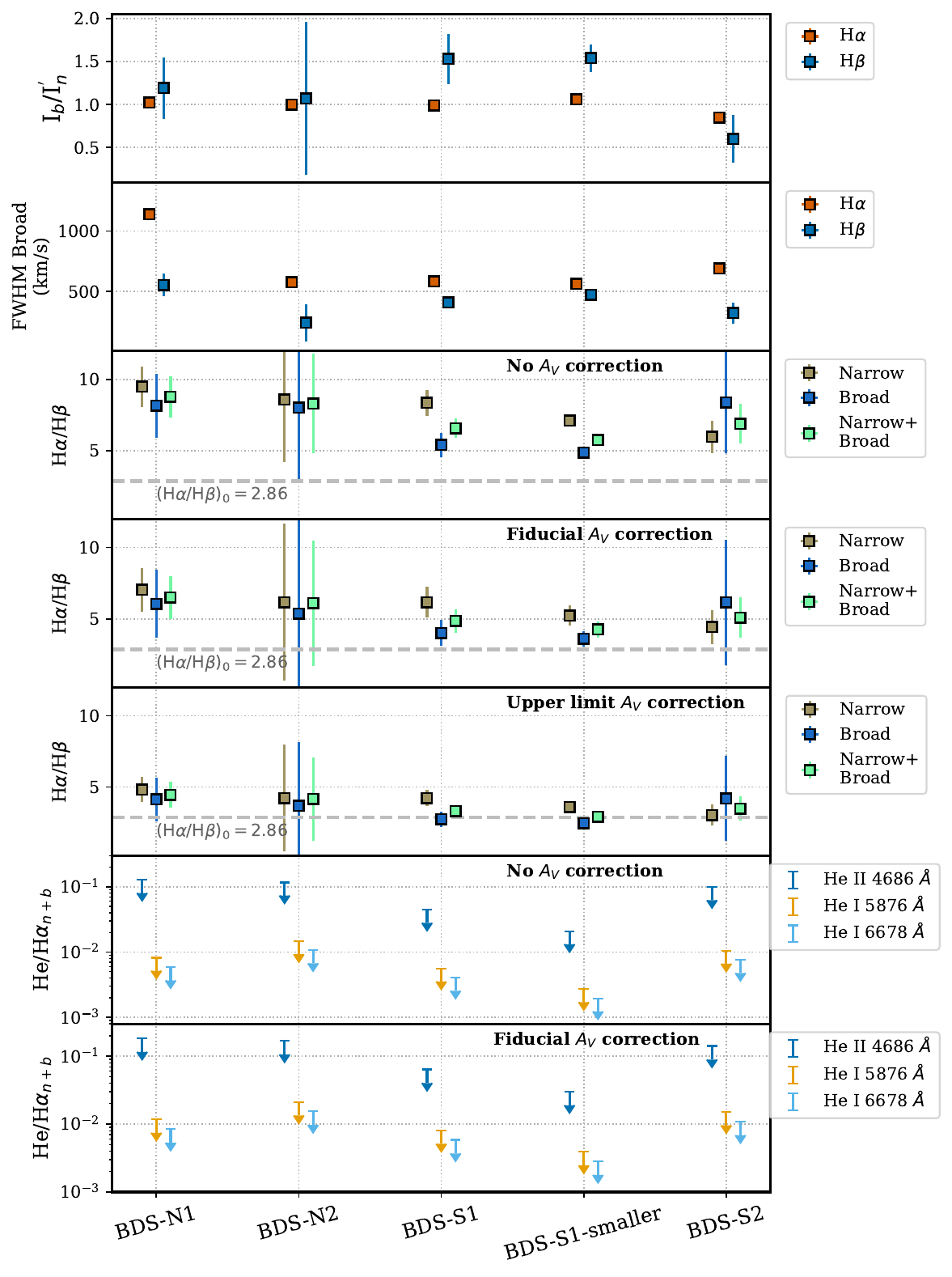}
    \caption{Properties of selected spectral lines of integrated Balmer-dominated shock regions from Figure \ref{fig:stacked_regions_positions}. From top to bottom -- (\emph{Row 1}): Broad-to-narrow intensity ratios of \ha and \hb lines, with the narrow flux corrected for radiative contamination. 
    (\emph{Row 2}): FWHM of the broad components of \ha and \hb. 
    (\emph{Row 3}): \ha to \hb ratio for the narrow, broad and narrow+broad components (colored), assuming no line of sight extinction.  
    (\emph{Row 4}): Same as Row 4, but with extinction-correction from Av-north and Av-south (Figure \ref{fig:stacked_regions_positions}), as explained in Section \ref{sec:balmer-decrement}. 
    (\emph{Row 5}): Same as Row 5, but with the maximum extinction-correction Av-rad (Figure \ref{fig:stacked_regions_positions}), as explained in Section \ref{sec:balmer-decrement}. 
    (\emph{Row 6}): 2$\sigma$-upper limit to the ratio of different He lines (wavelengths in the legend) to the narrow+broad \ha component (narrow has radiative contribution removed), assuming no extinction-correction. See Section \ref{sec:helines} for details.    (\emph{Row 7}): Same as Row 6, but with fiducial extinction correction from Av-north and Av-south as in Row 4.}
    \label{fig:stacked_regions_hahbprops}
\end{figure*}

The variation in line-profiles is further shown in Figure \ref{fig:stacked_regions_hahbprops}. We find that the \ibin for \ha is slightly lower than that of \hb. This is clearly seen in the southern shocks (BDS-S1, S1-smaller). In contrast, the \ibin for \ha and \hb in northern shock are similar within error-bars, while the effect is opposite in BDS-S2. The higher \ibin values for \hb are consistent with theoretical expectations of \lyg having lower conversion efficiency to \hb than \lyb to \ha \citep{Ghavamian2001,Heng07a}. The FWHM of broad \hb appears to be generally smaller than the \ha component (2nd row, Figure \ref{fig:stacked_regions_hahbprops}), though the gap significantly widens for the fainter northern and BDS-S2 regions. Wider broad components (especially in \hb) tend to be fainter, and thus lower signal-to-noise, so it is possible for the broad intensities to be under-estimated. 

We note the consistency between our measurements in BDS-S1-smaller and those in \cite{Ghavamian2001} in the same region (their Table 4) -- our \ha FWHM = 564.92$\pm$6.22 \kms, compared to their measurement of 562$\pm$18 \kms. Our \ha \ibin values are slightly lower (even after radiative shock correction) with 1.06$\pm$0.02, compared to their 1.18$\pm$0.03. The \hb \ibin ratio is also remarkably similar, with our 1.54$\pm$0.16, compared to their 1.54$\pm$0.17.

Before investigating the \ha/\hb ratios, we need to correct for extinction. While in principle this can be done with the \ha/\hb itself, the degeneracy with potential Lyman-line trapping prevents us from doing this fiber-by-fiber along the Balmer-dominated shocks. We therefore measure extinction in some by-eye selected fibers close to the Balmer-dominated shock fibers (but not including them). We pick two fiducial groups of fibers in the north and south (shown in Figure \ref{fig:stacked_regions_positions}) for extinction calculations, dubbed `Av-north' and `Av-south'. The \ha/\hb ratios in these two regions are $4.27\pm0.57$ and $3.83\pm0.36$ respectively, giving $A_V=1.21\pm0.41$ and $A_V=0.90\pm0.34$ mag respectively\footnote{Calculated with the \texttt{pyneb} software \citep{Luridiana2015}, assuming $R_V=3.1$ and a \cite{ODonnell94} extinction law}. We will apply the extinction in Av-north to BDS-N1 and BDS-N2, and Av-south to BDS-S1, BDS-S1-smaller, and BDS-S2. We are implicitly assuming that the \ha and \hb emission in Av-north and Av-south are also related to the SNR, and thus obscured by a similar dust column that is obscuring the Balmer-dominated shocks, which may not strictly be true. As a result, we also obtain $A_V$ from the shocked radiative cloud in the southwest (dubbed `Av-rad'), which we know for certain is related to the SNR. The caveat here now is that this cloud, likely being a dense cloudlet of interstellar or circumstellar origin, may have its own internal dust contributing to extinction in excess of the dust obscuring the non-radiative shocks. Indeed, the \ha/\hb ratio in the radiative cloud is $5.63\pm0.25$, corresponding to $A_V=2 \pm 0.13$. Since the Balmer-dominated shocks are not expanding into material as dense as the cloud (otherwise they would not be non-radiative), it is safe to assume the extinction along their line of sight is  $<2$ mag. We therefore use the $A_V$ in Av-north and Av-south as the fiducial extinction for the northern and southern Balmer-dominated shocks, and $A_V \approx 2$ in the radiative shock cloud as the upper limit.

The \ha/\hb results are shown in rows 3-5 in Figure \ref{fig:stacked_regions_hahbprops}, one for each assumption about extinction. The uncertainties in the ratios are propagated from both uncertainties in the line intensities, and uncertainties in the measured $A_V$. The uncertainties are typically larger in the northern fibers than the southern, driven by the generally fainter lines in the north. In all three instances of extinction correction, the \ha/\hb of the narrow components are generally higher than the broad components, and the narrow+broad \ha/\hb is intermediate (BDS-S2 seems to be an exception, but we note that the ratios for each component are consistent within their uncertainties). With extinction correction, the \ha/\hb values of all components decrease, with the lowest values obtained for the $A_V$ upper limit of 2 from Av-rad. 

The key result here is that the narrow \ha/\hb, even with the highest extinction correction, is still above the atomic physics limit of 2.86, while the broad component is consistent with the limit within error-bars. This is an observational support for Lyman-line trapping occuring in these collisionally-excited lines, with excess \lyb trapping (over \lyg) leading to a greater enhancement of \ha over \hb. The values of the narrow \ha/\hb in the range of 3-5 is also consistent with predictions for a range of \lyb optical depths \citep[e.g][]{CKR80,Tseliakhovich2012}. The results also confirm that the Lyman-line trapping mainly affects the cold neutrals, while the hot neutrals producing the broad components are unaffected (or little-affected) due to their large Doppler velocities. Our findings are consistent with results seen in other SNRs \citep[e.g.][]{Ghavamian2001, Ghavamian2017}.


\subsection{He-lines and Pre-shock Ionization} \label{sec:helines}
Another group of emission lines expected in Balmer-dominated shocks are those of He \citep[e.g.][]{Hartigan1999, Ghavamian2001, Ghavamian2002}, which is the next most abundant atom in neutral gas. Collisional excitation of He atoms passing through the shock can produce narrow lines similar to the Balmer-series before becoming fully ionized (e.g \ion{He}{1}$\lambda\lambda$5876, 6678). Singly-ionized He can be collisionally-excited to produce \ion{He}{2}$\lambda\lambda$ 4686, but with a unique broad-only signature, since it is an ion and thus, unlike H atoms, gets heated by the post-shock magnetic turbulence, broadening its distribution function. Evidence of \ion{He}{1} 6678 and marginal detection of \ion{He}{2}$\lambda\lambda$ 4686 was reported in SN1006 by \cite{Ghavamian2002}. The low-margin detections are expected due to the low He-abundance compared to H (He/H$\approx$0.1). The \ion{He}{2}:4686 broad line had been particularly proposed as a useful feature to study non-radiative shocks in fully-ionized gas, where Balmer-series will be absent \citep{Hartigan1999}. 

Importantly, the He line ratios with Balmer-lines can serve as unique diagnostics of the pre-shock neutral H and He fractions. \cite{Laming1996} gives the number of \heii 1640 photons per H atom going through the shock of 0.0012 for a 1000 \kms shock.  The CHIANTI database\footnote{\href{CHIANTI}{https://www.chiantidatabase.org/}}\citep{DelZanna2021} values of the \heii 4686/1640 ratio then imply 0.0016 \heii 4686 photons per H atom going through the shock (assuming that all the preshock He is He I or He II). Assuming 0.2 H$\alpha$ photons are produced per neutral H atom swept up by the shock \citep{Raymond1991}, we can express the intensity ratios of \heii to \ha as,
\begin{equation} \label{eq:neutralfraction}
\dfrac{I_{\mathrm{He},4686}}{I_{\mathrm{H}\alpha}} = \dfrac{0.008}{f_{H^0}}
\end{equation}
where $f_{H^0}$ is the neutral H fraction at the shock. 

We therefore looked for \ion{He}{2} lines in our five stacked regions (BDS-N1, BDS-N2, BDS-S1, BDS-S1-smaller, BDS-S2), and the results can be seen in Figure \ref{fig:stacked_regions_spectra}. As expected, the He lines are difficult to make out from the general noise in the spectrum. Running the Gaussian line-fitting at the location of these lines yield poor results (e.g negative or 0 amplitudes). Some cases however yield marginal detections. For example, for \ion{He}{2}\xspace4686, our line-fitter finds a $\sim$2.6$\sigma$ detection (in amplitude) in BDS-S1-smaller, with FWHM=$308\pm135$ km s$^{-1}$, and intensity of $(3.6\pm2.1) \times 10^{-15}$ erg s$^{-1}$ cm$^{-2}$. It would not be surprising if this was genuine as BDS-S1-smaller has bright and well-detected Balmer-components. The ratio of \ion{He}{2}:4686 to the narrow+broad \ha would then be $\left(9_{-5}^{+7}\right) \times 10^{-3}$, implying a neutral fraction $f_{H^0}=0.88_{-0.39}^{+1.11}$. With the fiducial extinction correction from Av-south, the ratio becomes $\left(1.3_{-0.7}^{+1}\right) \times 10^{-2}$, or $f_{H^0}=0.62^{+0.71}_{-0.27}$. 
Over the larger BDS-S1 region, the line-fitter yields a detection of \ion{He}{2}$\lambda\lambda$ 4686 of even lower significance ($\sim$2.2$\sigma$) and narrower line ($147\pm78$ \kms). The extinction-corrected neutral fraction in this case is $0.77_{-0.38}^{+1.61}$. 

However, given the low-significance of these detections, which is also visually apparent from the noisy spectra around these lines in Figure \ref{fig:stacked_regions_spectra}, we also treat the possibility of all five regions being non-detections of \ion{He}{2}$\lambda\lambda$ 4686, and instead show the 2$\sigma$ upper limits in the bottom two rows of Figure \ref{fig:stacked_regions_hahbprops}. To calculate the upper limit on the \ion{He}{2}$\lambda\lambda$ 4686, we assume that its amplitude has the same median and uncertainty as the noise-level on either side of the line, and its FWHM (+ uncertainties) is the same as the \ha line. We then create a large (10$^5$) mock sample of He, narrow \ha and broad \ha measurements from their measurement median and uncertainties, and report the 95th percentile of the distribution of the ratio $I_{\mathrm{He}}/(I_{\mathrm{H}\alpha,n}+I_{\mathrm{H}\alpha,b})$ in Figure \ref{fig:stacked_regions_hahbprops}. The $A_V$-corrected upper limit of 0.03 in BDS-S1-smaller is consistent with the 2.6$\sigma$ measurement, giving a lower-limit to the neutral fraction of 27\%. The upper-limits for the other regions are not as constraining, with values of $\sim$0.1-0.2, consistent with lower-limits of 0.04-0.08 on the neutral fraction. 

Overall, the measurements make a strong case that the southern shock is encountering a preshock medium with a higher ($\gtrsim$30\%) neutral fraction. Atomic clouds traced by the HI 21cm line are indeed visible in images of RCW86 towards the southwestern shock \citep{Sano2017}. Our measured neutral fractions and their limits are consistent with the neutral fraction expected, given the EUV radiation from the
nonradiative shock. Equation \ref{eq:neutralfraction} ignores any \ha emission from the shock
precursor \citep{Morlino2012},
which is extremely uncertain, but could increase the \ha by
10\% to 40\%. Taking that into account, the upper lower limit on
the neutral fraction is about 60-70\%, consistent with our findings.

We briefly mention here results for the neutral He lines. The \ion{He}{1}$\lambda\lambda$5876 line is particularly dubious as the continuum around the region is poorly-subtracted (Figure \ref{fig:stacked_regions_hahbprops}, this is near the red/blue break in the LVM spectrum, although un-subtracted foreground \ion{Na}{1} absorption lines could also be affecting). The \ion{He}{1}$\lambda\lambda$6678 is collisionally excited, so it is expected to be brighter than the recombination-driven \ion{He}{1}$\lambda\lambda$5876 in Balmer-dominated shocks \citep{Ghavamian2002}. While both lines appear undetected compared to the noise, we do find a 3.6$\sigma$ detection of \ion{He}{1}$\lambda\lambda$6678 in BDS-S1-smaller (bottom row, 4th panel, Figure \ref{fig:stacked_regions_spectra}). The line is narrow (about 52 \kms) and offset by $\sim$2\AA ($\sim$90 \kms) from its rest-wavelength. The line is sensitive to the electron temperature $T_e$, with \ion{He}{1}/H$\alpha$$<$1\% for log $(T_e/\mathrm{K})$$>$5. If we treat the line as associated with the BDS-S1-smaller shock, we measure a \ion{He}{1}$\lambda\lambda$6678 to H$\alpha$ (narrow+broad) ratio of 0.003, indicating generally high electron temperatures in the southern part of the shock. The 2$\sigma$ upper limits in all the regions also indicate \ion{He}{1}-to-H$\alpha$ ratios being $\lesssim$1\% (Figure \ref{fig:stacked_regions_hahbprops}).

\subsection{Intermediate \ha and Neutral Precursors} \label{sec:intermediateha}
The stacked regions created above also provide an opportunity to look for intermediate-velocity components of the Balmer lines, with widths of $\sim$few$\times$100 \kms. The presence of these components has been predicted by several papers \citep{Wagner2009, Morlino2012} due to the neutral return flux i.e. the escape of fast-neutrals, produced by charge exchange, to the shock-upstream. The velocity differential between these escaping neutrals and pre-shock ions heat the precursor, creating a secondary population of neutrals with distribution functions intermediate between the narrow and broad \ha-emitting population. The width of this intermediate component is proportional to the temperature of the pre-shock region heated by the neutral precursor, as opposed to the colder temperatures in the far-upstream reflected in the narrow comonent. These intermediate components were first claimed in LMC SNRs by \cite{Smith1994}, and subsequently in Tycho \citep{Ghavamian2000, Knevzevic2017} and N103B with IFU data \citep{Ghavamian2017}.  No such exploration in RCW86 has taken place, to our knowledge.
\begin{figure*}
    \centering
    \includegraphics[width=0.8\textwidth]{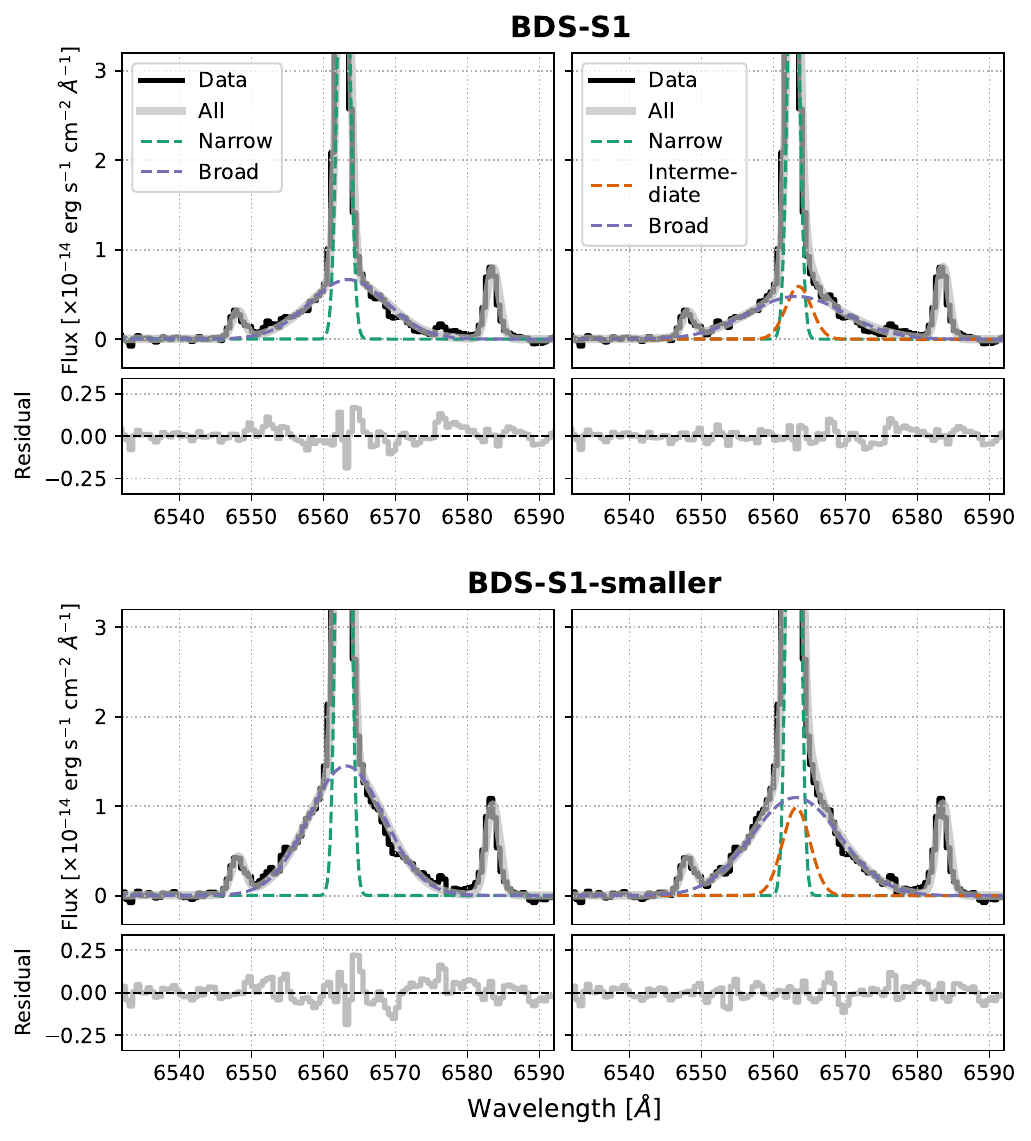}
    \caption{Fitting the southern shocks BDS-S1 (top) and BDS-S1-smaller (bottom) with an additional intermediate \ha component (Section \ref{sec:intermediateha}). Left panels shows the fits with a model that does not include an intermediate component, while right panel includes an intermediate component. Narrow, intermediate, broad components are colored as indicated. Grey denotes the sum of the fits (the \nii components are not separately shown, but they are part of the total fit). In each panel, the residual (data-model) is shown in the bottom. }
    \label{fig:intermediateha}
\end{figure*}

We redo the spectral fitting of the five Balmer-dominated shock filaments with a broad+intermediate+narrow \ha model. Only the southern BDS-S1 shock and its subset region BDS-S1-smaller appear to show evidence of a third intermediate component in Figure \ref{fig:intermediateha}. For BDS-S1, this third component has a FWHM of $192\pm15$ \kms, while the narrow and broad components have intensities and FWHM of $81\pm1$ and $725\pm24$ \kms respectively. The third component is clearly intermediate in width, to the narrow and broad components. The goodness of fit shows evidence of improvement based on the lower residuals in the model containing the intermediate component (right) versus the model without the intermediate component (left).  The $\Delta$AIC between intermediate and no-intermediate model (calculated in the same way as Eq \ref{eq:aic}) is 143, further supporting a better model fit. The evidence is stronger for BDS-S1-smaller, with $\Delta$AIC=210 supporting an intermediate component. The FWHM of the intermediate component is $207\pm12$ \kms, and that of the narrow and broad components are $80\pm0.5$ \kms and $663\pm12$ \kms respectively. Note that the broad component FWHM in the model including an intermediate component is higher than the models without them reported before (FWHM=$584\pm12$ \kms for BDS-S1 and $565\pm6$ \kms for BDS-S1-smaller). 

 The other regions -- BDS-S2, BDS-N1 and BDS-N2 -- were not successfully fitted with an added intermediate components. For BDS-S2 and BDS-N1, addition of an intermediate component leads to a fit where the broad component amplitude is forced to be negative. For BDS-N2, the broad component fit has amplitude=0. 

\begin{figure*}
    \centering
    \includegraphics[width=\textwidth]{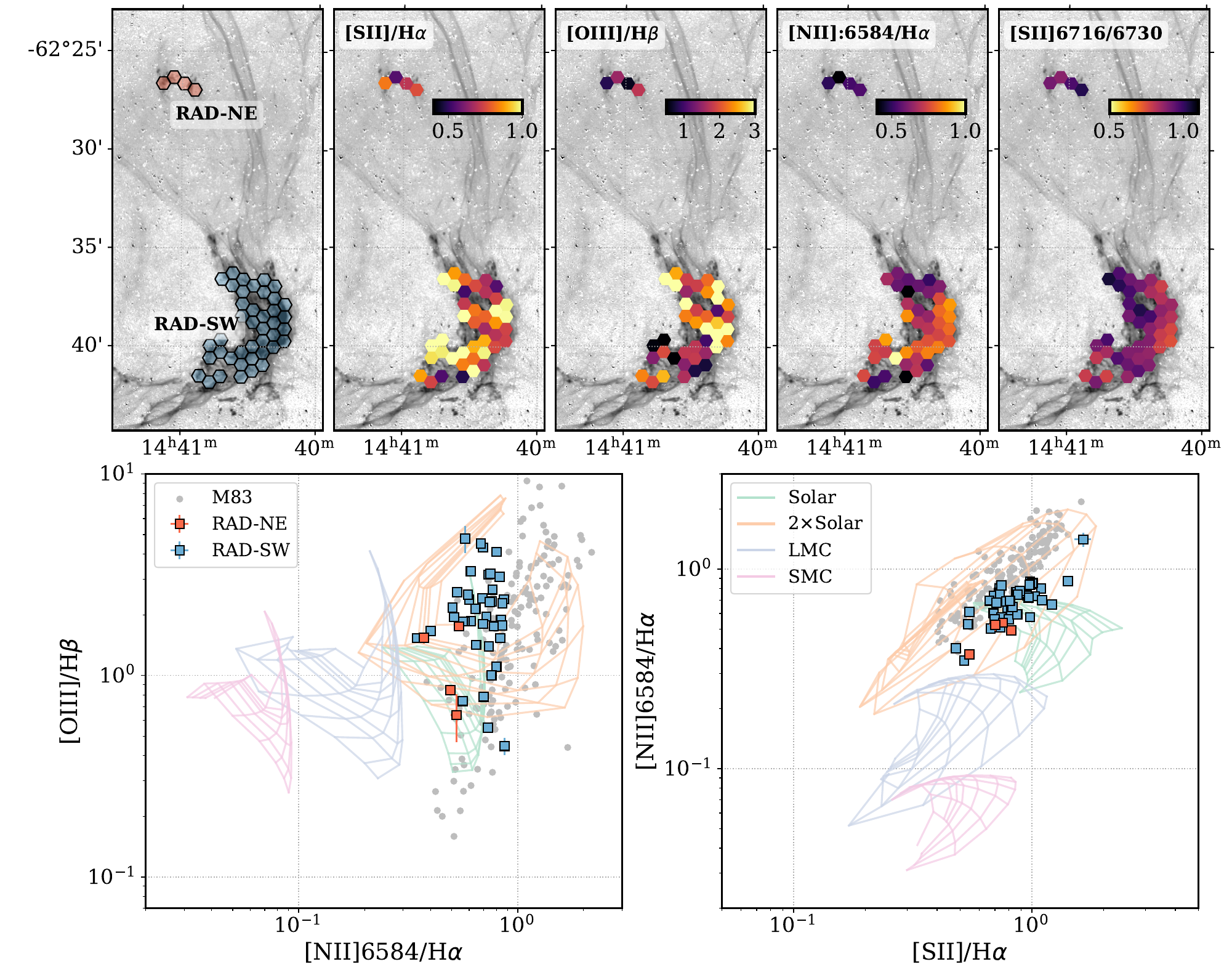}
    \caption{Spectral line properties of prominent radiative shocks in the southwest (``RAD-SW'') and northeast (``RAD-NE'') of RCW86. Top row shows the spatial locations of fibers in RAD-SW and RAD-NE, and subsequent panels show line ratios of \siiha, \oiiihb, \niiha, and \ion{S}{2}:6716/6731. Bottom rows shows the different BPT diagrams of the SNR. Left panel shows [\ion{O}{3}]/H$\beta$$-$\niiha BPT diagram of RAD-NE and RAD-SW, and Right panel shows \niiha$-$\siiha. Both middle and right bottom panels show integrated line-ratios of SNRs in M83 identified by MUSE spectroscopy from \cite{Long2022}. Both panels also show the MAPPINGSIII grid of radiative shock model predictions overlaid for four different abundances in \cite{Allen2008}. The model grid displayed covers the velocity range of 100-400 \kms, pre-shock density of 1 cm$^{-3}$, and pre-shock magnetic fields of 0.5-5 $\mu$G.} 
    \label{fig:radiativeshocks}
\end{figure*}

The above observations indicate that the southern Balmer-dominated shock has a substantial neutral return flux to produce detectable intermediate \ha emission. This is the first time that an intermediate \ha component has been claimed in RCW86. Similar analyses were used to argue for the presence of intermediate components of FWHM $\sim$ 144-218 \kms in N103B with IFU spectroscopy by \cite{Ghavamian2017}, and 180 \kms in Tycho \citep{Ghavamian2017, Knevzevic2017}. The broad components of these SNRs however are much wider than RCW86, with 1300-2300 \kms in N103B and 1800-2100 \kms in Tycho Knot g. Shock models of \cite{Morlino2012} including the neutral return flux were only calculated down to $v_s=10^3$ \kms as the distribution functions of the intermediate neutrals start to become negligible. The broad FWHM of BDS-S1-smaller implies $v_s=685-941$ \kms for \tetp=0.1-1 (Figure \ref{fig:fwhm-vs-VA08}), so we can atleast compare with the lower-limit of $v_s=1000$ \kms of \cite{Morlino2012}. A cursory comparison of our measured intermediate \ha FWHM=207 \kms with their Figure 10 for $v_s=1000$ \kms appears to rule out shocks where electrons and ions are fully-equilibrated both upstream and downstream, as well as shocks with no equilibration at all in the downstream (i.e. \tetp$\propto m_e/m_p$). Values of \tetp=0.01-0.1 might be able to explain the observed values. An extrapolation of the models below 1000 \kms, as well as inclusion of cosmic-rays, would allow to discern such scenario.

\subsection{Connection to Progenitors of Type Ia SNe}
Among the admittedly small sample of Galactic and Magellanic SNRs with secure Type Ia classifications, RCW 86 is unique in that it shows clear evidence for a fast, sustained outflow from the SN progenitor \citep{Badenes2007, Williams2011, Broersen2014, HMR2018}. The large spread of shock velocities that we find here further indicates that there is a large density contrast between the north and the south, which could be due to either large asymmetries in the ambient medium on scales of several pc, or an intrinsic anisotropy in the progenitor outflow. Moreover, \cite{Woods2017} and \cite{Woods2018} showed that the presence of neutral material ahead of the forward shock in Balmer-dominated Type Ia SNRs like RCW 86 rules out hot and luminous progenitors that would have emitted enough hard photons to ionize Hydrogen at the distance of the forward shock. This is remarkable in the case of RCW 86 because the accreting WD systems that are capable of ejecting fast outflows \citep{Hachisu1996, Han2004, Wolf2013, Cuneo2023} are often bright sources of X-ray and UV photons \citep{Mukai2017}. At least in the case of the progenitor of RCW 86, it seems the fast outflows were not accompanied by a large ionizing flux.

\section{An aside: Radiative Shocks in RCW86} 
\label{sec:radiativeshock}
Although our main focus is on the Balmer-dominated shock front, we briefly present the spatial and spectral properties of two distinct radiative shocked clouds in the LVM field: the large southwestern cloud (RAD-SW) and the two smaller northeastern cloudlets (RAD-NE) (Figure \textbf{\ref{fig:radiativeshocks}}). We defer a detailed analysis to future work and simply compare the data with predictions from radiative shock models. We select fibers belonging to these shocked clouds by visual inspection, deliberately avoiding the fibers classified as Balmer-dominated shocks with \nii/\ha$\leq$0.3 in Section \ref{subsec:tetp}. We focus on the widely-used bright forbidden lines, namely -- \ \niilu, \siilu, and \oiii$\lambda\lambda$5008. The fluxes for these lines are similarly fit as the Balmer-lines with Gaussian models with an underlying linear continuum (Section \ref{sec:methods}). For the \ha and \hb lines in these radiative shock regions, we use only the single-component fit (``Narrow-only'') results, as by definition these regions do not have broad components. No extinction correction is applied, as we only examine line-ratios that are closely-spaced in wavelength.

The top row of Figure \ref{fig:radiativeshocks} shows the spatial distribution of the radiative shocks, along with four commonly-used line-diagnostics for shocks -- [\ion{N}{2}]/H$\alpha$, \siiha, \oiiihb and \sii:6716/6731 ratio. The RAD-NE cloud generally has lower line-ratios than the RAD-SW cloud. The variable conditions in the cloud were already apparent from the inhomogenous structure seen in the narrowband image (Figure \ref{fig:lvmptg}), but also seen here in the wide range of values from the line-ratio maps. Some hint of spatial correlation can be seen between \sii/\ha and \nii/\ha, with both being lower in the lower and upper edges of the cloud where the density presumably falls off. The \nii/\ha however shows elevated structure at the leading (convex) edge of the cloud, where the clouds are likely being most compressed, while nothing similar appears in the \sii/\ha image. It is interesting however, to note the anti-correlation between the \oiii/\hb and \sii/\ha, with peaks in \oiii/\hb often occuring in fibers with low \sii/\ha and vice versa. Both clouds appear to have a relatively narrow range of \sii:6716/6731 doublet ratios. RAD-SW has \sii doublet ratios of 0.88$\pm$0.08, corresponding to $n_e=1125_{-292}^{+412}$ cm$^{-3}$ \citep[estimated using \texttt{pyneb},][]{Luridiana2015}. The shocked cloud appears to have a density gradient, increasing outwards in the direction of the shock. The RAD-NE cloud has lower densities, with \sii doublet ratio of $0.94\pm0.05$, corresponding to $n_e = 913_{-175}^{+220}$ cm$^{-3}$. 

The bottom panel of Figure \ref{fig:radiativeshocks} compares the BPT diagram of RCW86 RAD-SW and RAD-NE clouds with predictions from MAPPINGS-III suite of radiative shock models \citep{Allen2008}, along with an example comparison with extragalactic SNRs (as an example, we use M83 SNRs obtained from MUSE IFU spectroscopy by \cite{Long2022}). The MAPPINGS-III models assume planar shocks with velocities 100-1000 \kms (though we only show models up to 400 \kms) expanding into ambient medium with a pre-defined density, magnetic field, and abundance. These models assume the formation of complete shocks, with well-defined post-shock ionization, non-equilibrium, photoionization and recombination zones in steady-state where emission lines from atoms in various ionization states are produced. We show these models for four different abundances as defined in \cite{Allen2008}: SMC-like, LMC-like, solar \citep[based on ][]{Dopita2005}, and twice solar metallicity (see Table 1 of \cite{Allen2008} for details). Since MAPPINGS-III, there has been substantial updates to the atomic database and cooling functions used in the code \citep{Sutherland2017, Sutherland2018}, though for this paper, the publicly-available MAPPINGS-III library is sufficient for first comparison.

The comparison between our data, MAPPINGSIII models and the M83 SNRs leads to some interesting insights:
\begin{itemize}[leftmargin=*]
    \item Line ratios in RCW86 are generally in good agreement with the models, with [\ion{N}{2}]/H$\alpha$ values suggesting solar or super-solar abundances. This is consistent with \emph{Chandra} X-ray studies of the southwestern cloud \citep{Williams2011}
    \item M83 SNRs tend to show higher [\ion{N}{2}]/H$\alpha$ ratios than RCW86, which is expected given that M83 is a more metal-rich galaxy with a mix of core-collapse and Type Ia SNRs. M83 SNRs lie on the 2$\times$Z$_{\odot}$ MAPPINGS grid (with pre-shock magnetic fields of $\sim$1–5 $\mu$G), while RCW86 aligns with the same grid for $<$1 $\mu$G.
    \item Although [\ion{O}{3}]/H$\beta$ values fall within the model predictions, they are not perfectly reproduced by the MAPPINGS-III grid for shock-only models over the observed range of [\ion{N}{2}]/H$\alpha$.
    \item Nearly all fibers exhibit \ion{S}{2}/H$\alpha > 0.4$, the threshold typically used to distinguish shocks from photoionized gas.
\end{itemize}

In summary, the observed line-ratios capture the dispersion in shock velocities, densities, abundances, and magnetization within ths shocked clouds. A more detailed analysis of the full spectrum \citep[e.g][]{Dopita2019}, with updated atomic libraries \citep{Sutherland2017, Sutherland2018}, together with advanced simulations \citep{Makarenko2023, Guo2024, Smirnova2025, Romano2025} is warranted. Nevertheless, Figure \ref{fig:radiativeshocks} offers encouraging first results, demonstrating a good match between sub-pc scale observations of radiative shocks and planar shock model predictions, as well as consistency with integrated IFU measurements of extragalactic SNRs in M83. Currently, optical line-ratios yield the largest samples of extragalactic SNRs in galaxies, and offer the best way to differentiate SNRs from similar nebulae like HII regions and diffuse ionized gas, particularly for fainter objects \citep[e.g][]{Blair1997,Lee2014,Long2019,Moumen2019,CidFernandes2021,Long2022,Congiu2023,Winkler2023,Li2024,Kravtsov2024,DuartePuertas2024,Kopsacheili2024,Belfiore2024,Caldwell2025,Bracci2025}. The ongoing LVM survey will continue mapping Galactic and Magellanic SNRs, providing an empirical template to better distinguish SNRs from HII regions and diffuse ionized gas in extragalactic systems.

\section{Summary and Conclusion}
We present the first detailed ($\sim$0.3 pc scale) IFU survey of the Balmer-dominated supernova remnant (SNR) RCW86 with the SDSS-V Local Volume Mapper survey, with the goal of characterizing spatial variations of its Balmer-dominated non-radiative forward shock. The survey obtained optical spectra between 3600-9500 \AA\ of the southwestern portion of the SNR containing prominent radiative and non-radiative filaments (Figure \ref{fig:lvmptg}) in 1801 fibers of 35.3\arcsec\ apertures, covering a 30\arcmin wide hexagonal field of view. Four exposures of 900s duration were observed, reduced and stacked (Section \ref{sec:obs}). A set of background fibers were chosen to remove emission along the line of sight but unrelated to the SNR (Appendix \ref{sec:backsub}). The primary line-complex of interest -- \ha, \niilu -- were fitted with a multi-component gaussian model on a fiber-by-fiber basis, with an underlying linear continuum over a small wavelength range (Section \ref{sec:methods}). The \ha lines in particular were fitted with two models -- a Narrow-only model, which assumes the \ha has a single narrow component, and a Narrow+Broad model, which assumes the \ha has single and broad components. Fibers were selected to have a broad \ha component if they passed a certain goodness-of-fit measure (we use the Akaike Information Criteria). Fibers with negative amplitudes of any component, or with broad \ha components that are too narrow or too broad were also rejected. We also fitted major forbidden lines, such as \siilu, \niilu, and \oiii 5008, with single gaussian models for a brief analysis of the radiative shocked clouds (Section \ref{sec:radiativeshock}).   

Based on this, we produced the first map of Balmer-dominated shocks of RCW86 using spectroscopy, identifying \ha lines with a statistically significant broad component on a fiber-by-fiber basis (Figure \ref{fig:ifumaps}, \ref{fig:bdsfront}).  We then used the fibers with broad \ha components to explore a variety of topics related to collisionless shocks propagating in partially-ionized media. Our main results (details in Section \ref{sec:results} and \ref{sec:radiativeshock}) are as follows,
\begin{itemize}[leftmargin=*]
\item \textbf{Mapping the Balmer-dominated shock front:} The distribution of broad \ha fibers traces out the thin \ha filaments seen in optical narrow-band images (Figure \ref{fig:ifumaps}), tracing the non-radiative forward shock. The Balmer-dominated shock front is also visible in \hb, but deficient in forbidden lines such as \nii. 

\item \textbf{Shock velocity gradient:} A north-to-south gradient in shock velocities is revealed by the broad \ha FWHM, with speeds of 500-900 \kms in the south, to 1000-1500 \kms in the north (Figure \ref{fig:ifu-fwhm-ibin}). This indicates that the shock is encountering denser material in the south.  

\item \textbf{Electron-proton equilibration:} We confirm that electron and ion temperatures are not equilibrated in faster shocks (Figure \ref{fig:ibin-vs-VA08}), in line with previous results, based on the comparison of the observed broad-to-narrow ratios and FWHMs with collisionless shock models (ignoring precursors, however). Our IFU data enabled us to infer this result with a \emph{single} SNR owing to coverage of a wide velocity range ($\sim$200-2000 \kms) across the southwestern shock front.

\item \textbf{Broad component offset:} The broad component centroids appear typically redshifted relative to the narrow component by up to 100 \kms (Figure \ref{fig:shift-broad-narrow}), suggesting bulk post-shock motion away from the observer, or non-Maxwellian ion distributions. 

\item \textbf{Enhanced Balmer decrement:} Elevated H$\alpha$/H$\beta$ ratios ($\sim$3–5) in the narrow component, in contrast to broad components that are near the theoretical limit of 2.86 (Figure \ref{fig:stacked_regions_hahbprops}), indicate significant Lyman $\beta$ trapping in the cold neutral gas.

\item \textbf{Neutral fraction from He lines}: Marginal ($\sim2-2.6\sigma$) detections of \ion{He}{2}$\lambda\lambda$ 4686 in the stacked spectra along the southern shocks imply a high pre-shock neutral fraction (at least 30\%, consistent with up to 100\% within uncertainties).

\item \textbf{Intermediate \ha component:} Evidence for an intermediate \ha component with FWHM $\sim$192–207 \kms is found in the southern shock spectra (Figure \ref{fig:intermediateha}), consistent with predictions of neutral precursor models.\ This is the first confirmation of this phenomenon in RCW86, joining similarly observed phenomena in the Balmer-dominated SNRs Tycho, N103B and 0507-67.5. 

\item \textbf{Radiative shock analysis:} Though not our primary focus, a brief study of forbidden-line ratios in the radiative shocked clouds (Section \ref{sec:radiativeshock}, Figure \ref{fig:radiativeshocks}) indicates that the shocks are interacting with solar or slightly super-solar clouds with an outward density gradient, and with sub-pc scale line ratios matching planar shock model predictions from MAPPINGS, as well as integrated extragalactic SNR measurements.
\end{itemize}


We hope that this work encourages the further theoretical development of shocks, both Balmer-dominated and radiative, especially the impact of precursors formed due to cosmic rays and neutrals on the relevant observables such as \ibin and FWHM of \ha, \hb, He and other forbidden line properties. 

\begin{acknowledgments}

Funding for the Sloan Digital Sky Survey V has been provided by the Alfred P. Sloan Foundation, the Heising-Simons Foundation, the National Science Foundation, and the Participating Institutions. SDSS acknowledges support and resources from the Center for High-Performance Computing at the University of Utah. SDSS telescopes are located at Apache Point Observatory, funded by the Astrophysical Research Consortium and operated by New Mexico State University, and at Las Campanas Observatory, operated by the Carnegie Institution for Science. The SDSS web site is www.sdss.org.

SDSS is managed by the Astrophysical Research Consortium for the Participating Institutions of the SDSS Collaboration, including Caltech, The Carnegie Institution for Science, Chilean National Time Allocation Committee (CNTAC) ratified researchers, The Flatiron Institute, the Gotham Participation Group, Harvard University, Heidelberg University, The Johns Hopkins University, L’Ecole polytechnique fédérale de Lausanne (EPFL), Leibniz-Institut für Astrophysik Potsdam (AIP), Max-Planck-Institut für Astronomie (MPIA Heidelberg), The Flatiron Institute, Max-Planck-Institut für Extraterrestrische Physik (MPE), Nanjing University, National Astronomical Observatories of China (NAOC), New Mexico State University, The Ohio State University, Pennsylvania State University, Smithsonian Astrophysical Observatory, Space Telescope Science Institute (STScI), the Stellar Astrophysics Participation Group, Universidad Nacional Autónoma de México, University of Arizona, University of Colorado Boulder, University of Illinois at Urbana-Champaign, University of Toronto, University of Utah, University of Virginia, Yale University, and Yunnan University. J.G.F-T gratefully acknowledges the grants support provided by ANID Fondecyt Postdoc No. 3230001 (Sponsoring researcher), and from the Joint Committee ESO-Government of Chile under the agreement 2023 ORP 062/2023. National Optical Astronomy Observatory, which is operated by the Association of Universities for Research in Astronomy (AURA) under a cooperative agreement with the National Science Foundation. OE acknowledges funding from the Deutsche Forschungsgemeinschaft (DFG, German Research Foundation) -- project-ID 541068876. G.A.B. acknowledges the support from the ANID Basal project FB210003. EJJ acknowledges support by the ANID BASAL project  FB210003 and the FONDECYT Iniciaci\'on en investigaci\'on 2020 Project 11200263. KK gratefully acknowledges funding from the Deutsche Forschungsgemeinschaft (DFG, German Research Foundation) in the form of an Emmy Noether Research Group (grant number KR4598/2-1, PI Kreckel) and the European Research Council’s starting grant ERC StG-101077573 (“ISM-METALS"). 
\end{acknowledgments}

\begin{contribution}

SKS was responsible for carrying out the analysis, writing and submitting the manuscript.\ SKS, KSL, JCR, and RS formed the core paper team, and led the science interpretation.\ OVE, ARL, GAB, JDF and CB contributed to science interpretation.\ KSL, OVE, ND, EJJ, AMJ, IYK, KK, AMN, and SS play major roles in LVM operations, including setup and monitoring of observatory, scheduling, and pipeline development for data-reduction, analysis, sky-subtraction, and web-interfacing. All authors contributed to review and editing of manuscript. 

\end{contribution}

%
\facilities{SDSS-V LVM, CTIO/Schmidt}

\software{\texttt{astropy} \citep{Astropy}, \texttt{matplotlib} \citep{matplotlib}, \texttt{scipy} \citep{scipy}, \texttt{numpy} \citep{numpy}, \texttt{specutils} \citep{specutils}, \texttt{DS9} \citep{DS9} \texttt{regions} \citep{regions}
          }

\begin{appendix}

\section{Examples of Fibers failing broad \ha selection criteria} \label{sec:badspectra}
\begin{figure*}
    \centering
    \includegraphics[width=\textwidth]{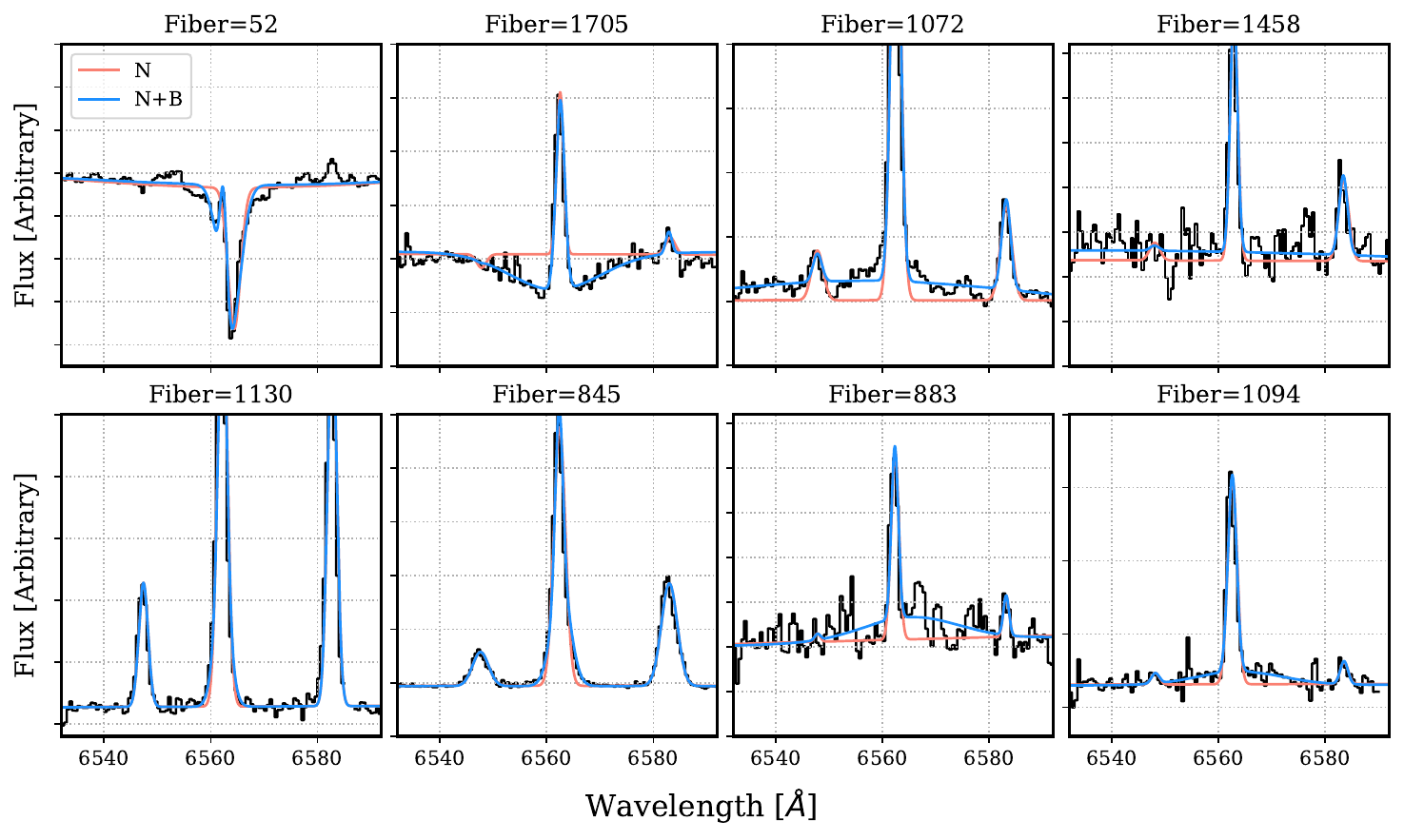}
    \caption{Examples of fibers that had a fitted broad \ha component, but were rejected based on the criteria in Section \ref{sec:methods}. The flux level of each spectrum was scaled by an arbitrary factor for viewing convenience (hence units are not shown). The label ``N'' refers to the Narrow-only model fit, and ``N+B'' refers to the Narrow+Broad model. See Appendix \ref{sec:badspectra} and Section \ref{sec:methods} for details.}
    \label{fig:badspectra}
\end{figure*}
Figure \ref{fig:badspectra} shows examples of fibers that were fitted a broad component but rejected based on one of more of the criteria mentioned in Section \ref{sec:methods}. As a reminder, 148 fibers passed our criteria for having a broad \ha component based on the Fiducial background subtraction (Appendix \ref{sec:backsub}) as shown in Figure \ref{fig:bdsfront}, and 1653 were rejected. The criteria used to reject are provided in parentheses. Most fibers are rejected by more than one criterion.
\begin{itemize}
\item Fibers 52 and 1705 failed the criterion (2): $A_i < 0$, i.e the fitted amplitudes of the \nii-\ha components were negative. Both are located on a prominent star in the narrowband image, so the broad component is likely fitting Balmer-absorption features of the stellar photosphere. About 60\% of the rejected fibers were rejected atleast due to this reason.
\item Fibers 1072 and 1458 failed the criterion (4): the ratio of the FWHM of the fitted broad component to the narrow component $>$20, i.e. the broad component is too broad. The FWHM of both fibers exceed 3000 \kms, which is unphysically large shock velocity in a single fiber given that most of the shock front has velocities three times lower. Both fibers appear to have a substantial broad component, but zooming-out the spectral axis reveals that there is substantial non-linear continuum (again likely from stars) dominating the fiber that forces a wide broad component. About 10\% of the rejected fibers are rejected atleast for this reason.
\item Fibers 1130 and 845 failed criterion (3): the ratio of the FWHM of the fitted broad component to the narrow component $<$3, i.e. the fitted component is too narrow. These typically occur near the radiative shocks, where the additional freedom of the Narrow + Broad model tends to overfit any wings beyond the central gaussian. The FWHM of the two fibers shown are 150-200 \kms. While the existence of broad components with these velocities may not be entirely unphysical, the spectra are too dominated by forbidden line emission and recombination to provide any useful insight on Balmer-dominated shocks. About 5.6\% of the fibers were rejected for this reason.
\item Fibers 883 and 1094 were rejected based on our $\Delta$AIC cutoff of 80 (criterion 1). The two fibers had $\Delta$AIC=55 and 77 respectively. The level of noise in the spectra of these fibers start to become substantial enough that the presence of a broad component becomes dubious. Although looking at the line profiles in Figure \ref{fig:badspectra}, one can definitely argue the existence of a broad \ha base, we chose to be conservative with the threshold in order to err on the side of purity (over completeness) of Balmer-dominated fibers for the analysis in Section \ref{sec:nonradiativeshock}. About 92\% of the rejected fibers fail this criterion.
\end{itemize}
The majority of the rejected fibers are therefore due to contamination by a strong stellar continuum (or a non-linear continuum that is improperly subtracted by our linear continuum model), and low signal-to-noise. A more careful fitting of the Narrow+Broad model with a stellar-library-based continuum-subtraction \citep[e.g., as done by the LVM data analysis pipeline in][]{Sanchez2025} will likely fix the issue, and will be attempted in future studies, though we avoided this for the present study owing to its being more computationally expensive. We note however that the simple fitting procedure in Section \ref{sec:methods} is still quite effective at excluding fibers where broad components are not expected, finding them mostly along the thin filaments tracing the forward shock (Figure \ref{fig:bdsfront}).  
\section{Local Background Subtraction} \label{sec:backsub}

The results we have presented depend on the accuracy with which we can derives of \ha\ from the filaments in RCW86, especially of the narrow component of \ha since the ``sky'' that is subtracted as part of the data reduction process described in  Section \ref{sec:obs} is taken from data obtained some degrees away from RCW86, and since there is a possibility that there is faint but detectable foreground or background emission from diffuse gas along the line of sight to RCW86.  To address this, we have subtracted a ``local background'', prior to analyzing the SNR spectra.  We do not anticipate that the sky varies within the field of view of LVM, we have to be concerned that the foreground or background emission does. Here we address our sensitivity to this problem.

\begin{figure*}
    \includegraphics[width=\textwidth]{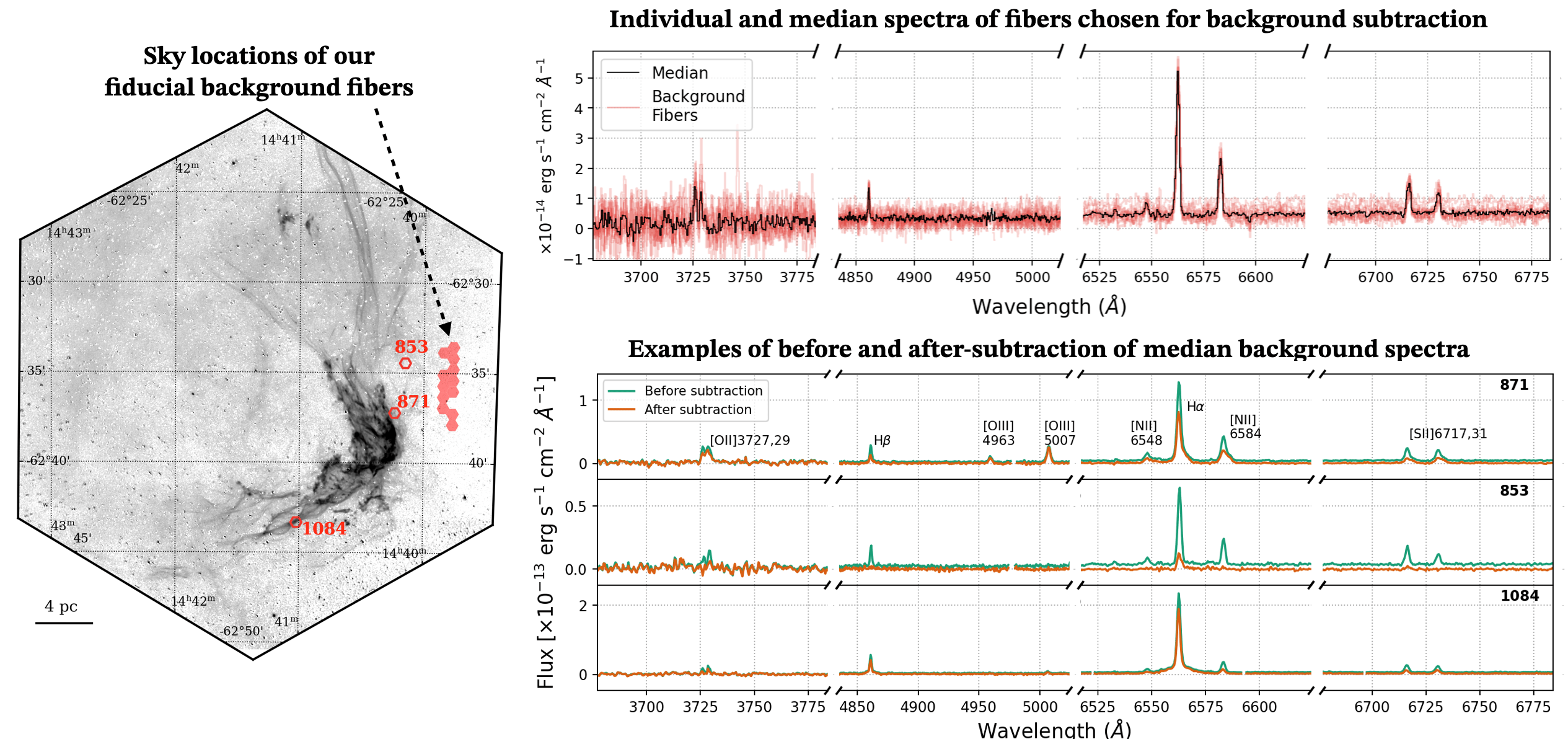}
    \caption{Demonstration of background fiber subtraction. \emph{Left panel} shows the LVM footprint, location of our fiducial background fibers (red-filled), and example fibers whose spectra we display on the right (red-hollow). \emph{Top right panel} shows portions of the spectral axis of the individual fibers (red) highlighting the bright lines (e.g \oii, \hb, \oiii, \ha, \nii, \sii). Black is the median spectra that we subtract from all fibers. \emph{Bottom panel} shows before- and after-background-subtraction spectra of the three selected red fibers from the left. The example fibers include one containing strong forbidden-line features (871), one located on empty sky (853) and one on a Balmer-dominated shock (1084).}
    \label{fig:backsub}
\end{figure*}

Figure \ref{fig:backsub} shows the location of the background fibers we use for the results in the main text (called ``Fiducial’’). These were selected by-eye to be outside the SNR-emitting region of RCW86, and west of the forward shock. We picked fibers that are contiguous, but at the same time not containing significantly elevated continuum levels (e.g. fibers coincident on bright stars). We then take a median of the spectra from these fibers (example shown for a single exposure on the top panel of Figure \ref{fig:backsub}), and subtract from the spectra of all fibers. The results are shown on the right panels of Figure \ref{fig:backsub}. We show three representative fibers here -- the top panel (fiber 853) located on blank sky, the middle panel (fiber 871) located on a radiative shock, and the lower panel (fiber 1084) located on a prominent Balmer-dominated filament. The subtraction results in an overall reduction in the continuum and line fluxes, but to different degrees. The lines in the blank sky fiber are almost completely from the background, and disappear after subtraction. In contrast, the lines in the radiative shock are primarily from the shock itself, and only decrease by a small amount after subtraction. The same result is seen for the Balmer-dominated filament, but it is noteworthy that the broad \ha component survives the subtraction, further supporting this feature being intrinsic to the shock, and not an artifact from the background.

\begin{figure*}
    \includegraphics[width=\textwidth]{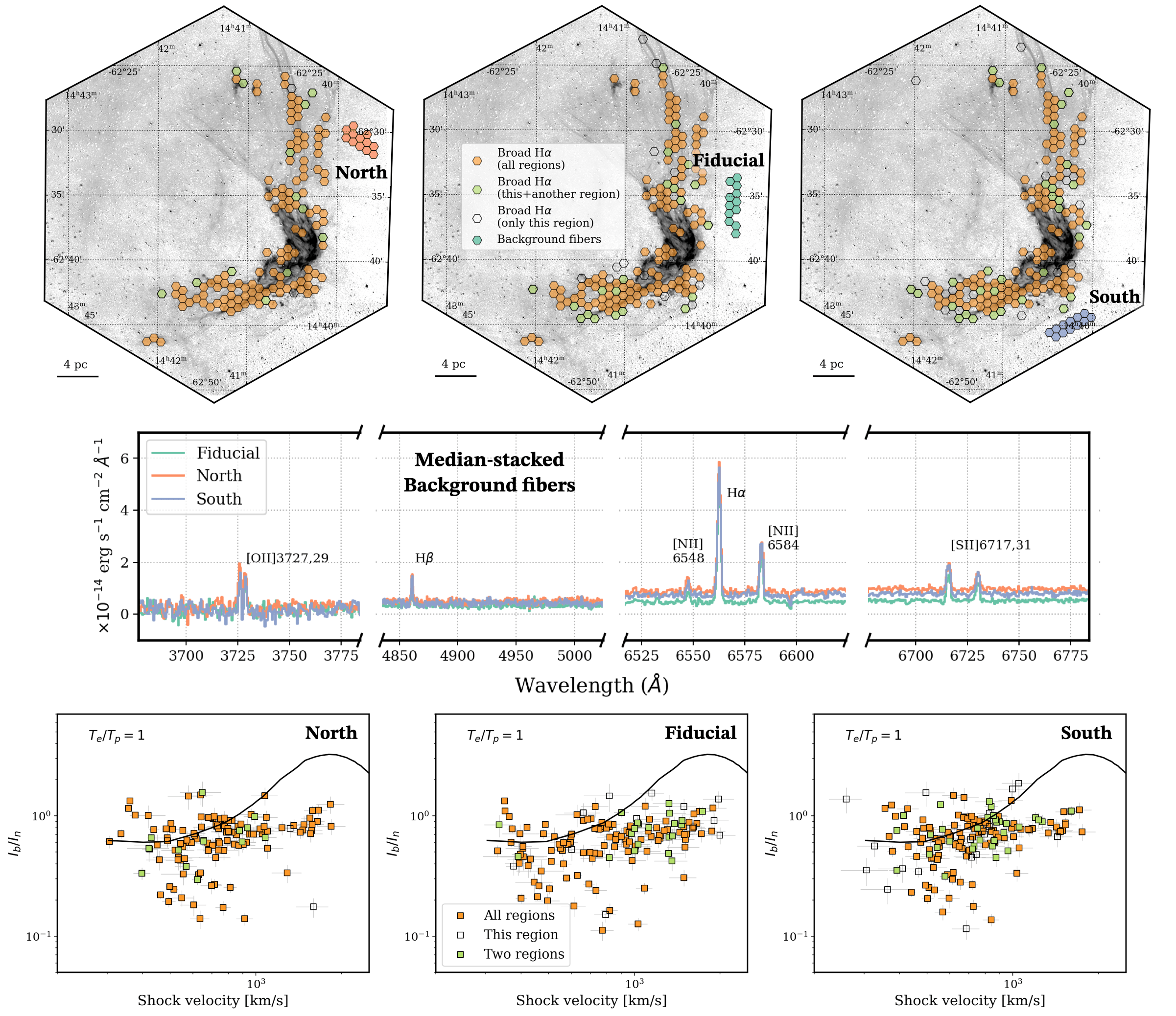}
    \caption{Effect of choice of background fiber locations on selection and science of broad \ha fibers. \textit{Top} row shows the three groups of fibers used for background subtraction -- North (red, left), Fiducial (middle, emerald), and South (right, violet) -- and the resulting fibers with broad \ha based on selection criteria in Section \ref{sec:methods} (hexagonal points). Orange fibers passed the broad \ha criteria for all three background fiber groups. Green fibers passed the broad \ha criteria for the background fibers shown in that panel \emph{and} at least another panel. Unfilled fibers passed the broad \ha criteria only for the background fibers shown in that panel. \emph{Middle} row shows the median spectra in the three background regions, sliced into wavelength ranges of interest. \emph{Bottom} row shows the analysis of \ibin vs shock-velocity for three different \tetp models (Section \ref{subsec:tetp}), but repeated for the three different background cases. Orange, green and unfilled points have the same meaning as the top row.}
    \label{fig:backsub_exp}
\end{figure*}

How do our results change if we change the fibers being used for background? This is demonstrated in Figure \ref{fig:backsub_exp}. We focus the effect of different choice of background fibers on the selection of fibers with the broad \ha component only, since that forms the core science in this paper. Aside from Fiducial, we pick two regions in the northern (NORTH) and southern (SOUTH) pre-shock region of RCW86. The fibers are once again chosen to be contiguous, west of the forward shock and outside the SNR-emitting region, and without bright stellar emission. The median sky spectra of the different sky regions are shown in the top panel of Figure \ref{fig:backsub_exp}. The continuum varies by about a factor of two (measured between 6600 and 6700 \AA), with the fiducial sky region being the faintest, and the north sky region the brightest among the three. The difference increases from the blue to the red end of the spectrum. The varying background levels from the North to Fiducial to South can also be seen in the narrowband CTIO image in Figure \ref{fig:backsub}.

Despite the varying sky level, the selection of broad fibers is not as significantly impacted, as shown in the IFU maps (middle panels) in Figure \ref{fig:backsub_exp}. The Fiducial case yields 148 fibers that pass our broad-fiber criteria in Section \ref{sec:methods}, while the North and South cases yield 130 and 164 respectively. However, 115 fibers pass the broad \ha test for all three cases (shown in orange in Figure \ref{fig:backsub}), constituting 78\%, 88\% and 70\% of the broad fibers in each of the three cases -- Fiducial, North and South -- respectively. The spatial locations of these 115 fibers are also consistent with the large-scale, arc-shaped geometry of the forward shock of RCW86, and co-located with the majority of the prominent, thin, Balmer-dominated filaments seen in the CTIO image. In addition to the common fibers, each case has fibers that are classified as broad in at least two images (21 in Fiducial, 13 in North, and 32 in South). Thus, only 8\%, 1.5\% and 10\% of the fibers are classified as broad solely in the Fiducial, North and South cases.

The impact on the broad-to-narrow ratio vs shock velocity is also shown in Figure \ref{fig:backsub_exp}, bottom panel for the Fiducial, North and South cases. We only show the \tetp=1 model as illustration, but the result applies to the other models too. There are not many stark changes in the plots aside from the number density of points, and shifts in the parameter space for some fibers in each case. The range of \ibin and velocities spanned by the fibers are about the same in each case, and the main science result -- that the higher \tetp models are inconsistent with higher shock velocities still hold, as we see shocks $>$1000 \kms still being excluded by the model. 

In summary, the results on the Balmer-dominated science can be concluded as robust to the choice of fibers for background subtraction, with the three cases investigated here causing only about a 10-20\% difference in fibers that are classified as broad, with most of these appearing on the edges of the major Balmer-dominated filaments. The broad-to-narrow \ha vs velocity analysis in the context of electron-ion-equilibration models is robust as well. 

\end{appendix}

\bibliographystyle{aasjournalv7}
\bibliography{main_submitted}

\end{document}